\documentclass[review]{elsarticle}
\usepackage{lineno,hyperref}
\modulolinenumbers[5]

\journal{Journal of \LaTeX\ Templates}
\pdfoutput=1 

\usepackage{float}
\usepackage{color}

\def\beq#1{\begin{equation}\label{#1}}
\def\eeq{\end{equation}}
\def\beqa#1{\begin{eqnarray}\label{#1}}
\def\ee{\end{eqnarray}}
\def\be{\begin{eqnarray}}

\def\mycomment#1{\relax}
\def\beq{\begin{equation}}

\begin{document}

\begin{frontmatter}

\title{\boldmath PBH evaporation, baryon asymmetry, and dark matter }

\author{A. Chaudhuri and A. Dolgov\fnref{myfootnote}}
\address{Novosibirsk State University, Novosibirsk, Russia 630090}
\fntext[myfootnote]{ITEP,  Bol. Cheremushkinsaya ul., 25, 117218 Moscow, Russia}

\author[mymainaddress]{arnabchaudhuri.7@gmail.com}

\author[mysecondaryaddress]{dolgov@fe.infn.it}


\begin{abstract}
Sufficiently light primordial black holes (PBH) could evaporate in the very early universe and dilute the preexisting baryon 
asymmetry and/or the frozen density of stable relics. The effect is especially strong in the case that PBHs decayed  if and when 
they dominated the cosmological energy density. The size of the reduction is first calculated analytically under the simplifying 
assumption of the delta-function mass spectrum of PBH and in instant decay approximation. In the realistic case of exponential
decay and for an extended mass spectrum of PBH the calculations are made numerically. Resulting reduction of the frozen 
number density of the supersymmetric relics opens for them a wider window to become viable dark matter candidate.
\end{abstract}


\end{frontmatter}

\linenumbers

\section{Introduction \label{s-intro}}

Primordial black holes might be abundant in the early universe and even dominate  for a while the cosmological energy density.
In the latter case they would have an essential impact on the baryon asymmetry of the universe, on the fraction of dark matter particles, 
and  would lead to the rise of the density perturbations at relatively small scales. 

Usually primordial black holes (PBH) are supposed to be created by the Zel'dovich-Novikov  (ZN) mechanism~\cite{ZN-PBH} (see 
also~\cite{Carr-Hawking}). According to ZN, a PBH could be created, if the density fluctuation, $\delta \rho /\rho$, at the horizon size 
happened to be larger than unity. In this case this higher density region would be inside its own gravitational radius and became a
black hole. With the accepted 
Harrison-Zeldovich spectrum of primordial fluctuations~\cite{Hdelta,Zdelta} the process of PBH creation can result in a significant 
density of PBHs. 

The mass inside horizon at the radiation dominated (RD) stage of he universe evolution is equal to
\be
M_{hor} = m_{Pl}^2 t,
\label{M-hor}
\ee
where the Planck mass is $m_{Pl} \approx 2.176 \times 10^{-5}$ g and $t$ is the cosmological time (universe age). 
Thus the initial moment of the creation of PBH with mass $M$ can be taken as 
\be 
t_{in} (M)  = M/m_{Pl}^2 .
\label{t-in}
\ee

{It is mostly assumed that the mass spectrum of PBH created by ZN mechanism is very narrow.  It is usually 
taken in a power law form 
or even  as delta-function. There are, however, quite a few other scenarios of PBH formation. We can mention, in particular,
the mechanism suggested in ref.~\cite{AD-JS,DKK}, which leads to log-normal mass distribution and may, in principle, create
PBH with masses up to thousands and even millions solar masses due to production of the BH seeds during cosmological inflationary
stage. Other mechanisms of PBH production initiated at inflation are considered in refs.~\cite{INN,BLW}. Some more
work on PBH formation with extended mass spectrum include refs.~\cite{9,10,11,SMPBH}.}
The creation of PBH due
to a phase transition in the primeval plasma is studied in~\cite{RKS}. A recent review on massive PBH formation
can be found in~\cite{DER}. 

{The log-normal mass spectrum became quite popular during last few years, being employed for the description of massive
PBH observed in the present day universe. The analysis of chirp mass distribution of the LIGO events~\cite{KAP}
very well agrees with the log-normal mass spectrum.}

Here we consider much smaller PBH masses such that the  black holes evaporated early enough, well
before the Big Bang Nucleosynthesis (BBN). Because of calculational problems we take the PBH mass spectrum
either as a flat one bounded between some $M_{min}$ and   $M_{max}$ or a power law one, also bounded between
$M_{min}$ and   $M_{max}$, but continuously vanishihg at the boundaries.  
The latter spectrum can be quite close numerically to the  log-normal one.

Though such short-lived PBH decayed long before our time, their impact on the present day universe may be well noticeable.
Firstly, PBH decays could pour a significant amount of entropy into the primeval plasma and diminish the magnitude of earlier 
created baryon asymmetry or diminish the relative (with respect to the relic photon background) density of dark 
matter particles~\cite{DNN,ACh-AD}.
On the other hand, baryon asymmetry could be generated in PBH evaporation~\cite{YaBZ-bar, AD-bar}, and dark
matter could also be created in this process. We neglect however, the second kind of the processes and consider
only dilution of baryons and dark matter particles by the PBH evaporation. 
{Indeed it can be shown that the stable supersymmetric relics produced in the process of PBH evaporation make negligible 
contribution to the density of dark matter, see Appendix A.}

An interesting well known effect, not touched in this work, is the rise of density perturbations during early matter dominated stage.
If there existed an epoch of the early PBH domination, the rising density perturbations could create 
small scale clumps of matter in the present day universe such as globular clusters or even dwarf galaxies,

In the scenario, which is considered below, the universe is supposed to be initially in radiation dominated (RD) stage, i.e.
the cosmological matter at this stage mostly consisted of relativistic species.  
The cosmological energy density during this epoch was equal to
\be
\rho_{rel} ^{(1)}= \frac{ 3 m_{Pl}^2}{32 \pi t^2} .
\label{rho-rel}
\ee
and the scale factor at this epoch evolved as
\be 
a_{rel} (t) = a^{(in)} \, \left(\frac{t}{t_{in}}\right)^{1/2} .
\label{a-rel}
\ee

If sufficiently large density of PBH was created during this period and if PBH were massive enough to survive up to the moment 
when they became dominating in the universe, the cosmological expansion law turned into the non-relativistic one and the energy 
density started to tend asymptotically to:
\be
\rho_{nr} = \frac{ m_{Pl}^2}{6 \pi (t+ t_1) ^2} .
\label{rho-nr}
\ee
Ultimately all PBH evaporated producing relativistic matter and the expansion regime returned to the relativistic one:
\be
\rho_{rel}^{(2)} = \frac{ 3 m_{Pl}^2}{32 \pi( t+t_2)^2} .
\label{rho-rel2}
\ee

In thermal equilibrium the energy density of relativistic particles is equal to
\be 
\rho_{rel} = \frac{\pi^2 g_* (T) T^4}{30},
\label{rho-of-T}
\ee 
where  $T$ is the plasma temperature and $g_*(T)$ is the number of relativistic species in the plasma at temperature $T$.

It is known, see e.g.~\cite{DG-VR,CB-AD}, that in thermal equilibrium state of the cosmological plasma with zero chemical 
potentials  the entropy in the comoving volume is conserved:
\beq
 s = \frac{\rho+\cal{P}}{T} a^3=const,
 \label{1}
 \eeq
where $\rho$ is the energy density of the plasma and $\cal{P}$ is its pressure.

In usual baryogenesis scenarios non-conservation of baryonic number took place at very high temperatures, while at low 
temperatures baryon non-conservation was switched off. So at late cosmological epochs baryonic number density, $N_B$, was also
conserved in the comoving volume. Correspondingly the baryon asymmetry, i.e the ratio 
\be
\beta = N_B / s = const
\label{beta}
\ee
remained constant in the course of the universe expansion if there was no entropy influx into the plasma.

There are several realistic mechanisms of entropy production in the early universe. For example, entropy rose in the course of the
electroweak phase transition, even if it was second order (or mild crossover). The entropy rise could be at the 
level of 10\%~\cite{ACh-AD}. If  in the course of the cosmological evolution a first order phase transition took place, 
e.g.  the QCD one, the entropy rise can be gigantic. Some entropy rise could be created by 
{the residual annihilation of out-of-equilibrium of
non-relativistic dark matter particle after they practically} decoupled from the plasma (froze). 

In this work we consider a hypothetical case of the universe which
at some stage was dominated by PBHs and calculate the dilution of the
preexisting baryon asymmetry and a relative decrease of the number density of DM particles. 
{We show that in a reasonable scenario of PBH creation weakly interacting massive particles (WIMPs), 
denote them $X$, with the annihilation cross section $\sigma_{ann} v \approx \alpha^2/ m_X^2$, $\alpha \sim 10^{-2}$
may have masses somewhat larger than TeV, avoiding the LHC bound, and be realistic candidates for dark matter.}

{The parameter space of supersymmertry is known to be significantly restricted by LHC~\cite{LHC-lim}, 
but some types of the lightest supersymmetric particles  (LSP) still remain viable candidates for dark 
matter~\cite{LHC-DM1,LHC-DM2}. An excessive entropy release, discussed in this paper, can lead to a wider
class of possible dark matter LSPs.  }

The paper is organized as follows. In the next section we present a simple estimate of the entropy release for the case of delta-function
mass spectrum of PBHs, instant decay approximation for PBH, and instant change from the initial RD stage to MD stage and back.
In Sec. 3 the exact solutions for the cosmological evolution and the entropy release for the mixture of relativistic matter and decaying PBHs
with the delta-function mass spectrum 
are found. Sec. 4 is devoted to the study of the evolution for two examples of the extended mass spectrum. In sec. 5 we analyze the results 
and conclude.  {Appendix A  is devoted to calculations of the number density of $X$-particles directly produced by PBH decays, the
subject which is somewhat away from the main line of this paper.
In Appendix B the expressions of the analytically calculated integrals entering the evolution equations are presented.}

\section{Instant change of expansion regimes and instant evaporation   \label{sec-instant}}

We consider here the simplest model of PBHs with fixed mass $M_0$ with the number density at the moment of creation:
\be
\frac{dN_{BH}}{dM} = \mu^3_1 \,\delta (M-M_0),
\label{dN-dM-1}
\ee
where $\mu_1$ is a constant parameter with dimension of mass.

All these PBHs were created at the same moment $t_{in} (M_0) = M_0/m_{Pl}^2$, see eq. (\ref{t-in}). Assume that the fraction of the 
PBH energy (mass) density at production was:
\be
\frac{\rho_{BH}^{(in)}}{\rho_{rel}^{(in)}} = \epsilon \ll 1 
\label{epsilon}
\ee

If we disregard the PBH decay and if the interaction between PBH and relativistic matter
can be neglected, then both ingredients of the cosmic plasma evolve independently and so:
\be
\rho_{rel} (t)= \left(\frac{a^{(in)}}{a(t)}\right)^4 \rho_{rel}^{(in)},\,\,\,\,\,  
\rho_{BH} (t) = \left(\frac{a^{(in)}}{a(t)}\right)^3 \rho_{BH}^{(in)}
\label{rho-evol}
\ee

Let us consider the case when densities of relativistic and non-relativistic (PBH) matters became equal 
at $t = t_{eq}$, before the PBH decay. According to eqs.~(\ref{epsilon}) and (\ref{rho-evol}) it takes place when:
\be
\frac{\rho_{BH} (t_{eq})}{\rho_{rel} (t_{eq})} = \epsilon\, \frac{a(t_{eq})}{a_{in}} = 1 .
\label{equil}
\ee
We assume in this section that at $t < t_{eq}$ the universe expansion is described by purely relativistic law, when the scale 
factor evolves according to eq.~(\ref{a-rel}). Correspondingly we find 
\be
t_{eq}=t_{in}/\epsilon^2 .
\label{teq}
\ee
 PBHs would survive in the primeval plasma till equilibrium
if $t_{eq} - t_{in} < \tau_{BH}$, where the life-time of PBH with respect to evaporation is given by the expression~\cite{page76}:
\be
\tau (M) \approx 3\times 10^3 N_{eff}^{-1} M_{BH}^3 m_{Pl}^{-4}  \equiv C\,\frac{M_{BH}^3}{m_{Pl}^4},
\label{tau-BH}
\ee 
{where $C \approx 30$}, if  
the effective number of particle species with
masses smaller than the black hole temperature, is { $N_{eff} \approx 100$}.
{ (In reality $g_*$ is closer to 200, but this difference is not of much importance.)}
The black hole temperature is equal to:
\be
T_{BH} = { m_{Pl}^2 \over 8\pi M_{BH}} .
\label{Tbh}
\ee
Thus the condition that the RD/MD equality is reached prior to BH decay reads:
\be
M_{BH} > \left[ \frac{m_{Pl}^2}{C} \left( \frac{1}{\epsilon^2} -1 \right)\right]^{1/2} \approx  \frac{m_{Pl} }{\sqrt{C} \,\epsilon }.
\label{M-max}
\ee

According to the assumption of the instant change of the expansion regime, the scale factor  after the 
equilibrium moment is reached, i.e. for $t>t_{eq}$, started to evolve as
\be
a_{nr} (t) = a_{rel} (t_{eq}) \left(\frac{t+t_{eq}/3}{4t_{eq}/3}\right)^{2/3}
\label{a-nr}
\ee
and the cosmological energy density drops according to the non-relativistic expansion law:
\be
\rho_{BH} = \frac{m_{Pl}^2}{6\pi \,(t + t_{eq}/3)^2} .
\label{rho-BH}
\ee
Such forms of eqs.~(\ref{a-nr}) and (\ref{rho-BH}) are dictated by the continuity of the Hubble parameter and of  the 
energy density (i.e. by equality of $\rho_{rel}$ and $\rho_{BH}$) at $t=t_{eq}$.
Such a regime lasted till $t=\tau_{BH}$, when instant explosion of PBHs created new relativistic plasma with the temperature:
\be
T_{heat}^4 = \frac{5 m_{Pl}^2}{\pi^3 g_*(T_{heat}) (\tau_{BH} + t_{eq}/3)^2} .
\label{T-heat-4}
\ee
Instant thermalization is here assumed.

The temperature of the relativistic plasma coexisting with the dominant PBH dropped down as the scale factor:
\be
T_{rel} = T_{eq}\,\frac{ a_{eq}}{ a_{nr} (\tau) } = T_{eq}\,\left(\frac{4 t_{eq}}{3\tau_{BH} + t_{eq}}\right)^{2/3} .
\label{T-rel-drop}
\ee

Correspondingly the temperature of the newly created by the PBH decay relativistic plasma could be much higher than 
$ T_{rel}$ given by eq.~(\ref{T-rel-drop}).
The entropy suppression factor, which is equal to the cube of the ratio of the temperatures of the new 
relativistic plasma created by the PBH instant evaporation  to temperature of the  "old" one, 
plus unity from the entropy of the old relativistic plasma is equal to:: 
\be
S = 1 + \left( \frac{T_{heat}}{T_{rel}}\right)^3 = 
1 + \left(\frac{a(\tau_{BH})}{a_{eq}}\right)^{3/4} = 
1 + \sqrt{\frac{3\tau_{BH}}{4 t_{eq}}}\,\left( 1 + \frac{t_{eq}}{3\tau_{BH} }\right)^{1/2}
\label{suppr-factor}
\ee
Our approach is valid for $\tau_{BH} \geq t_{eq}$ and in the limiting case of $\tau_{BH} = t_{eq}$
the entropy suppression factor is $S= 2$ coming from the
relativistic matter and from PBH in equal shares. Since the minimal value of 
\be
\frac{\tau_{BH} }{ t_{eq}} = \frac{C M_{BH}^2 \epsilon^2}{m_{Pl}^2}
\label{tau-to-teq}
\ee
is equal to unity, the minimal mass of PBH for which we can trust the approximate calculations presented above is
\be
M_{BH} > M_1^{min} \equiv \frac{m_{Pl} }{\epsilon\, \sqrt{C}} \approx 4\cdot 10^6 \,{\rm g} \left(\frac{10^{-12}}{\epsilon}\right) ,
\label{M-min-1}
\ee
where $C=30$, according to eq.~(\ref{tau-BH}).


\begin{figure}[h!]
\includegraphics[]{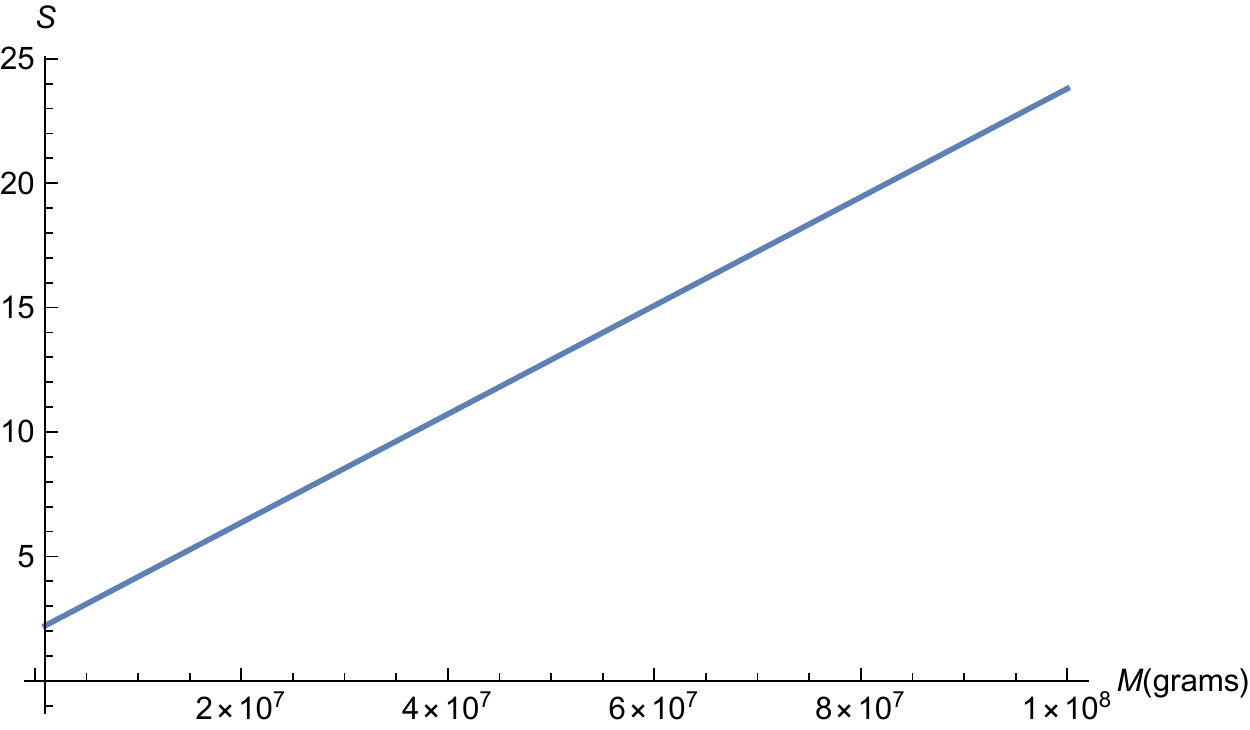}
\caption{Entropy suppression factor due to PBH decay 
in the instant decay approximation as a function of BH mass, starting from $M_1^{min}$, up to $M= 10^8 M_\odot$
for $\epsilon = 10^{-12}$. }
 
\label{f-entropy-1}
\end{figure} 

\begin{figure}[h!]
\includegraphics[]{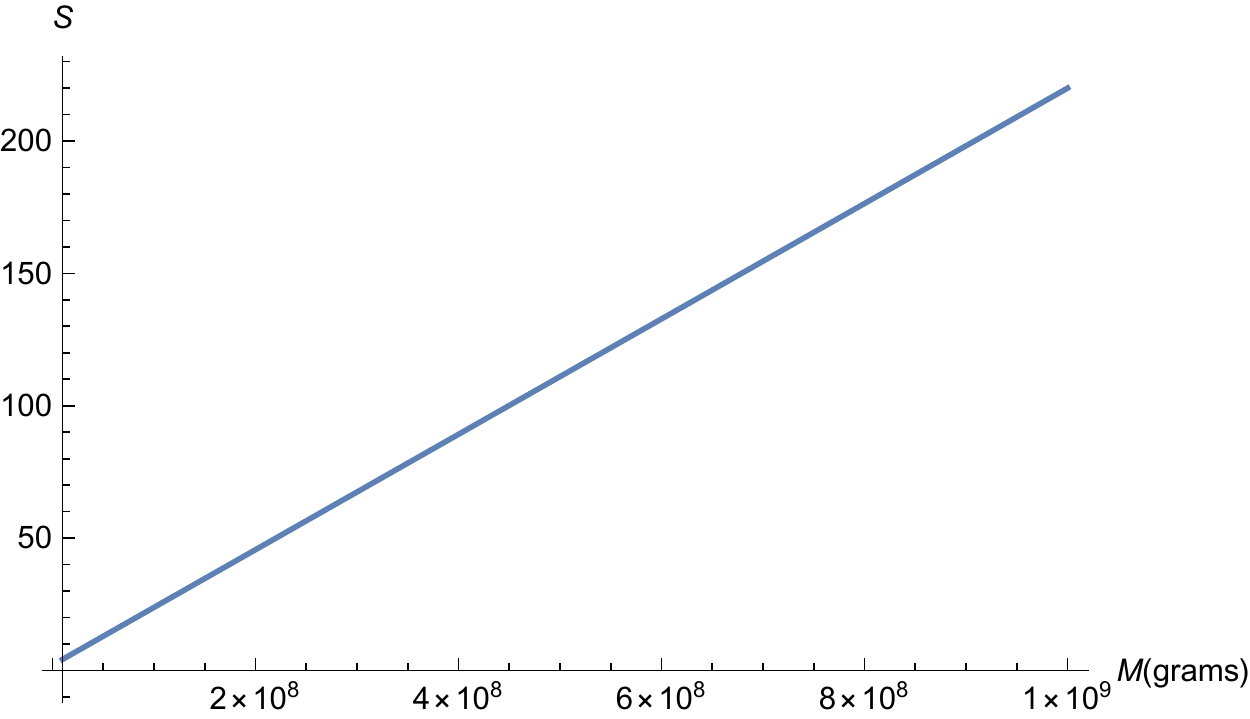}
\caption{Entropy suppression factor due to PBH decay 
in the instant decay approximation for larger masses up to maximal mass  $M =10^9 M_\odot$
as a function of BH mass for $\epsilon = 10^{-12}$. }
\label{f-entropy-2}
\end{figure} 

For large  $ \tau \gg t_{eq} $, when $S$ is large, it is approximately equal to
\be{
S\approx \sqrt{\frac{3\tau_{BH}}{4 t_{eq}}}\ = \frac{\sqrt{3C}\,\epsilon M}{2 m_{Pl}} = 
2.14 \cdot 10^{-7} \left(\epsilon/10^{-12}\right) \left(M /\rm{g}\right) .}
\label{suppress-2}
\ee
 
The PBH mass is bounded from above by the condition that the heating temperature after evaporation should be higher 
than the BBN temperature, $\sim 1 $ MeV. From eq.~(\ref{T-heat-4}) it follows that
\be
T_{heat} \approx 0.06 m_{Pl} \left(\frac{m_{Pl}}{M_{BH}}\right)^{3/2}.
\label{T-heat}
\ee
Hence the PBH masses should be below $10^9$g.  

The entropy suppression factors for $\epsilon = 10^{-12}$ as functions of $M_{BH}$ are  presented in Figs.  \ref{f-entropy-1} and
\ref{f-entropy-2} for small and large masses respectively.


\section{Exact solution for delta-function mass spectrum \label{s-exact}}

Here we relax the instant decay approximation and solve numerically equations describing evolution of the cosmological energy densities
of non-relativistic PBHs and relativistic matter. It is convenient to work in terms of dimensionless time variable  $\eta = t/\tau_{BH}$, when
the equations can be written as::
\be 
\frac{d \rho_{BH}}{d\eta} = - (3H\tau + 1) \rho_{BH} \label{dot-rho-BH} ,\\
\frac{d \rho_{rel}}{d\eta} = - 4H\tau \rho_{rel} +  \rho_{BH}.  \label{dot-rho-rel}
\ee

We present the energy densities of PBH and relativistic matter  respectively in the form:
\be
\rho_{BH} &=& \rho_{BH}^{(in)} \exp{(-\eta +\eta_{in})} y_{BH} (\eta)/ z(\eta)^{3}  ,
\label{rho-BH-sol} \\
\rho_{rel} &=& \rho_{rel}^{(in)}  y_{rel} (\eta) /z(\eta)^{4},
\label{rho-rel-sol}
\ee
where $y_{rel}^{(in)} = y_{BH}^{(in)} = 1$ and 
\be 
\eta_{in} = \frac{m_{Pl}^2}{C M^2_{BH} } \ll 1 .
\label{eta-in}
\ee
The constant $C$ is determined in Eq.~(\ref{tau-BH}).

The redshift factor $z(\eta) = a(\eta)/a_{in}$ satisfies the equation:
\be
\frac{dz}{d\eta} = H \tau_{BH}\,z ,
\label{dz-deta}
\ee
where the Hubble parameter $H$ is determined by the usual expression for the spatially flat universe:
\be
\frac{3 H^2 m_{Pl}^2}{8\pi} = \rho_{rel} + \rho_{BH}.
\label{H}
\ee
Using equations (\ref{rho-rel-sol}) and  (\ref{rho-BH-sol}) with
$\rho_{rel}^{(in)} $ given by Eq.~(\ref{rho-rel}) at $t= t_{in}$ and bearing in mind that 
$\rho_{BH}^{(in)} = \epsilon \rho_{rel}^{(in)} $ we find
\be
H \tau_{BH} = \frac{C}{2}\,\frac{M_{BH}^2}{m_{Pl}^2}\,\left( \frac{y_{rel}}{z^4} + \frac{\epsilon}{ z^3 e^{\eta-\eta_{in}} } \right)^{1/2} .
\label{H-tau}
\ee

Evidently Eq. (\ref{dot-rho-BH}) with $\rho_{BH} $ given by (\ref{rho-BH-sol}) is solved as
\be
y_{BH}(\eta) = y_{BH}^{(in)} =1,
\label{yBH-of-eta}
\ee
while $\rho_{rel} (\eta) $ satisfies the equation:
\be 
\frac{dy_{rel}}{d\eta} = \epsilon z (\eta) e^{-\eta + \eta_{in} }.
\label{dy-rel-deta}
\ee
Equations (\ref{dz-deta}) and (\ref{dy-rel-deta})
can be solved numerically with the initial conditions at $\eta = \eta_{in}$ 
\be
y_{bh}=y_{rel}=z=1 .
\label{yin}
\ee

\begin{figure}[h!]
\includegraphics[scale=0.7]{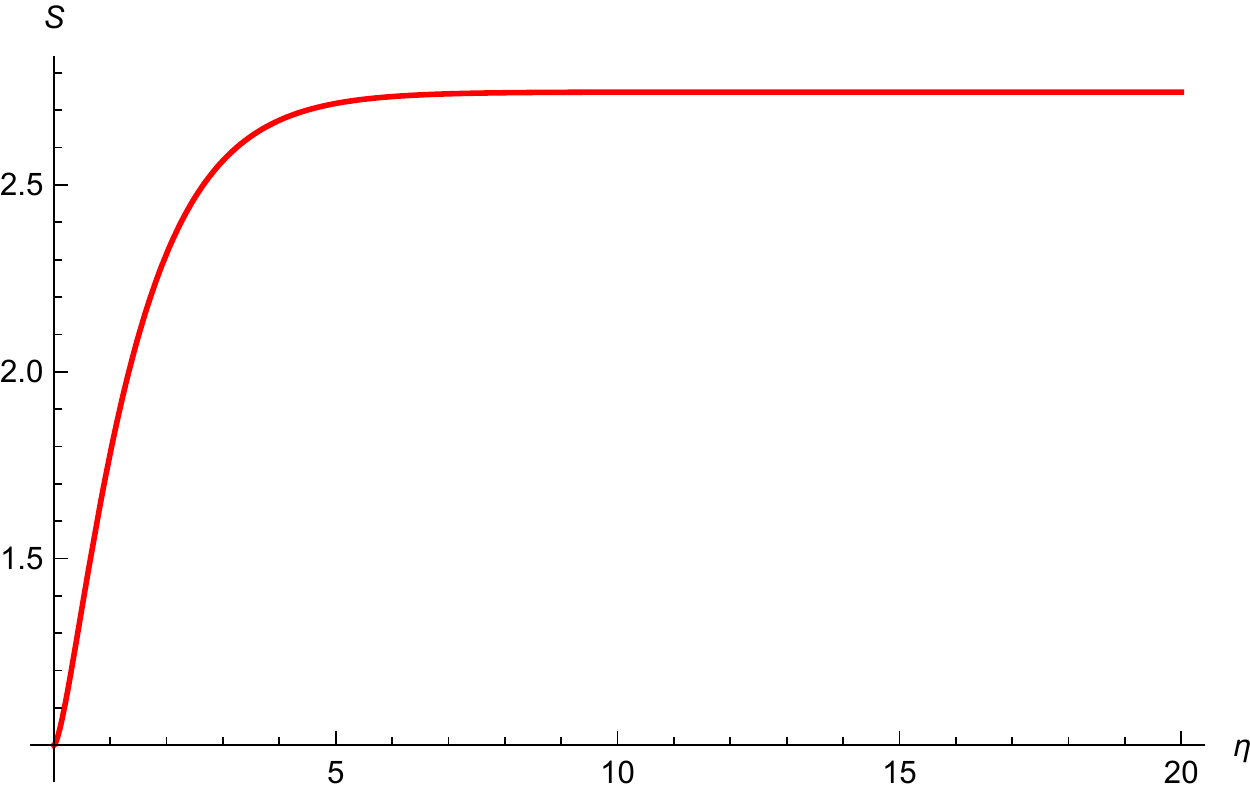}
\caption{The temporal evolution of $S$ for $M_{BH} = 10^7$ g and $\epsilon = 10^{-12}$}.
\label{s-of-eta-7}
\end{figure}

\begin{figure}[h!]
\includegraphics[scale=0.7]{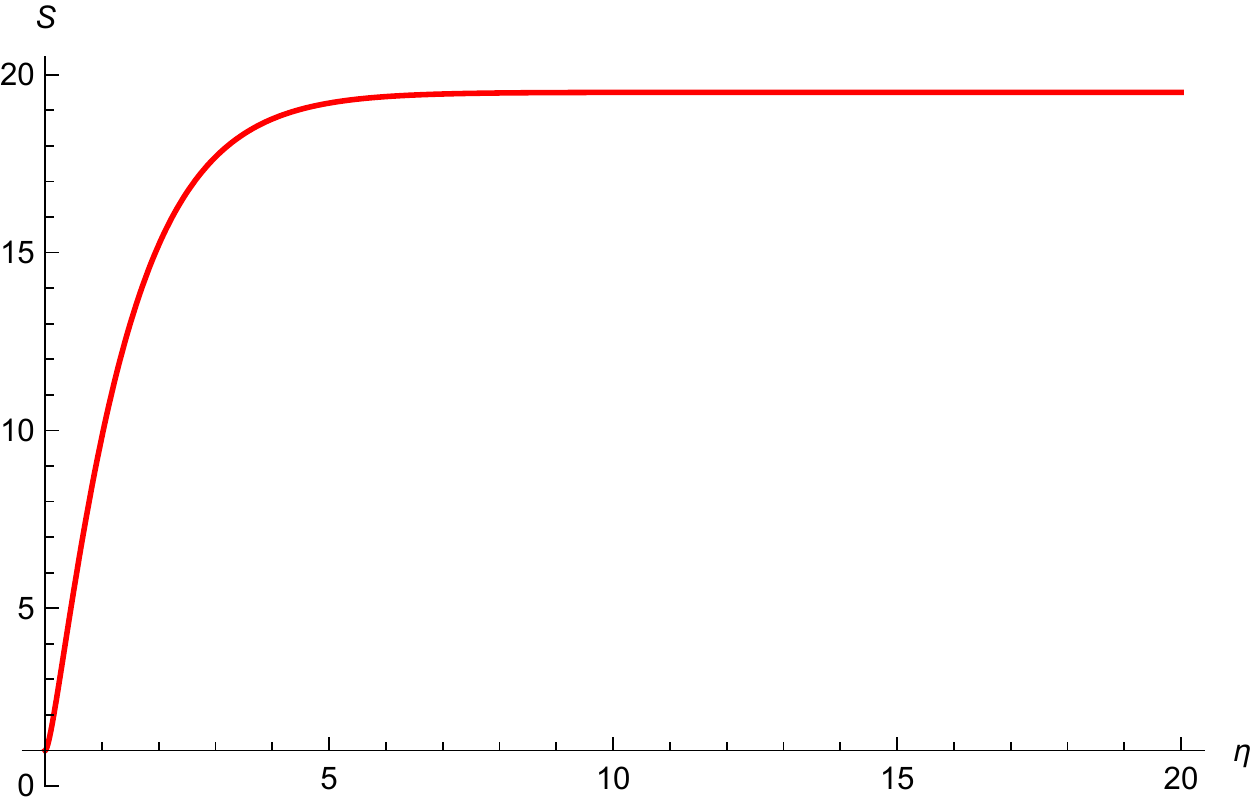}
\caption{The temporal evolution of $S$ for $M_{BH} = 10^8$ g and $\epsilon = 10^{-12}$}
\label{s-of-eta-8}
\end{figure}

\begin{figure}[h!]
\includegraphics[scale=0.7]{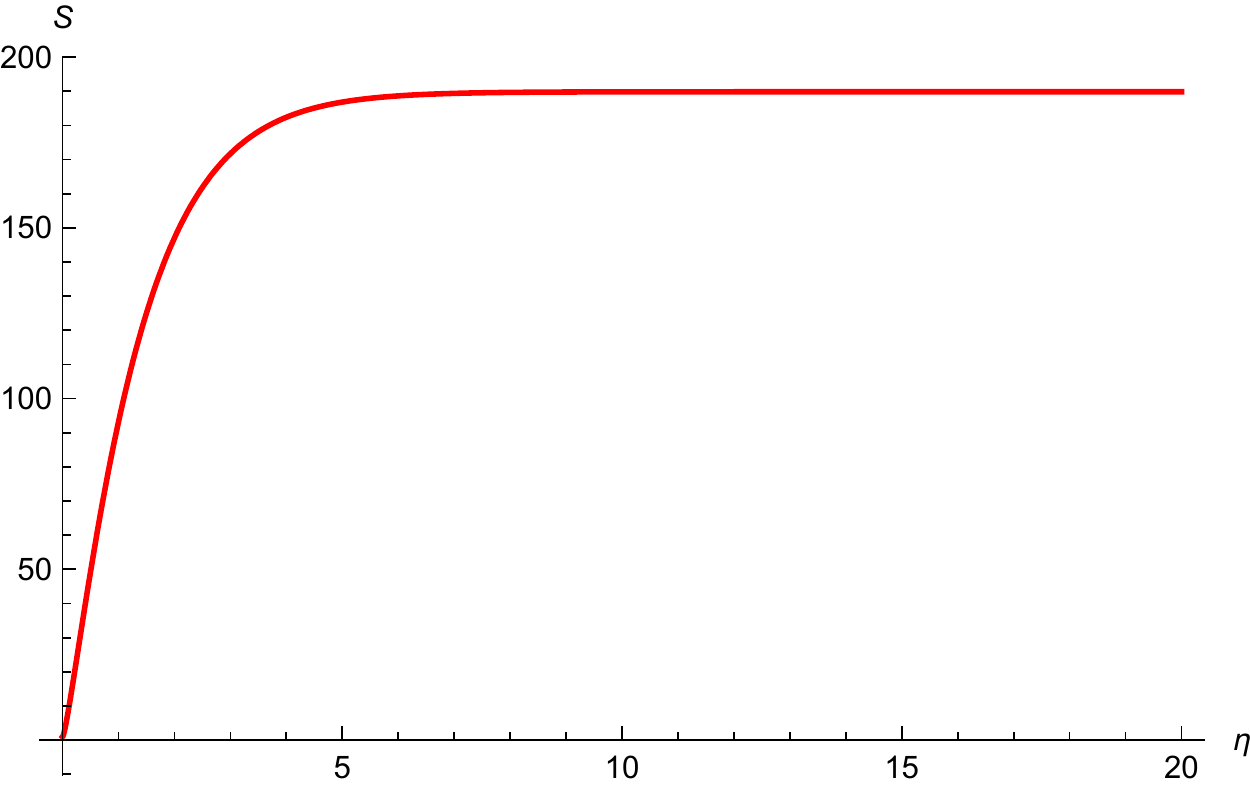}
\caption{The temporal evolution of $S$ for $M_{BH} = 10^9$ g and $\epsilon = 10^{-12}$}
\label{s-of-eta-9}
\end{figure}

However, a huge value of the coefficient  $H \tau$ makes the numerical procedure quite slow. To avoid that we introduce the new function 
$W$ according to:
\be 
z = \sqrt{W}/\epsilon 
\label{w}
\ee
and arrive to the equations:
\be 
\frac{d W}{d\eta} &=& C \epsilon^2 \left(\frac{M}{m_{Pl}}\right)^2 \left( y_{rel} + {\sqrt{W}}\,e^{-\eta + \eta_{in} }\right)^{1/2} ,
\label{dW}\\
\frac{d y_{rel}}{d\eta} &=& \sqrt{W} e^{-\eta + \eta_{in} },
\label{dy-rel}
\ee
where $W(\eta_{in}) =\epsilon^2 $. Entropy release from PBH evaporation can be calculated as follows. In the absence of PBHs 
the quantities conserved in the comoving volume evolved as $1/z^3$. With extra radiation coming from the PBH evaporation
the entropy evolves as $y_{rel}^{3/4} /z^3 $, see eq. (\ref{rho-rel-sol}). Hence the suppression of the relative number density of
frozen dark matter particles or earlier generated baryon asymmetry is equal to:
\be 
S = \left[ y_{rel} (\eta ) \right]^{3/4}
\label{S}
\ee
when time tends to infinity. The temporal evolution of $S$ is depicted in figs. \ref{s-of-eta-7},  \ref{s-of-eta-8},  \ref{s-of-eta-9},
for different values of 
$M_{BH}= 10^7, 10^8,10^9$ grams and $\epsilon = 10^{-12}$.

For large $\eta$  (in fact  $\eta > 15$) $S$ tends, as expected, to a constant value. The comparison of these figures 
with figs.~\ref{f-entropy-1} and \ref{f-entropy-2} demonstrates
perfect agreement between approximate calculations and the exact ones.

\begin{figure}[h!]
\includegraphics[scale=0.7]{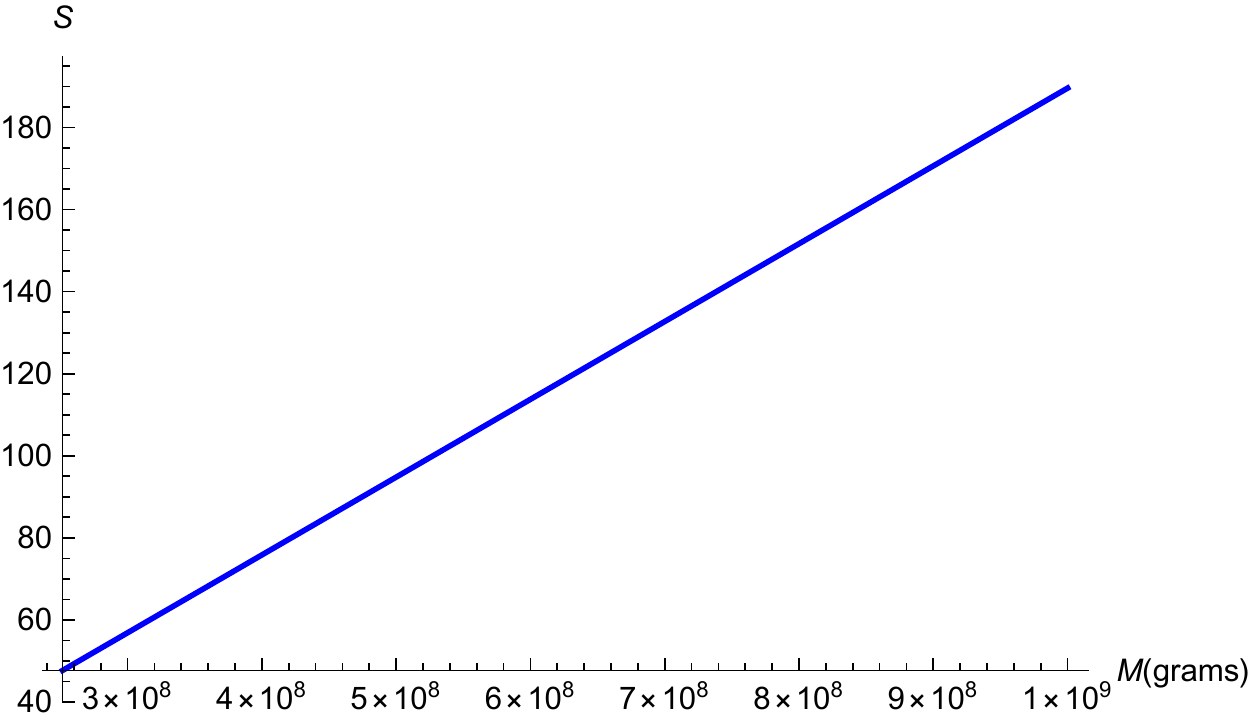}
\caption{The entropy suppression factor as a function of mass for $\epsilon = 10^{-12}$
}
\label{s-of-m}
\end{figure}

In fig.~\ref{s-of-m} the asymptotic value for the entropy suppression factor is presented as a function of PBH mass.
for $\eta = 10^{-12}$ in perfect agreement with approximate calculations depicted in figs.~\ref{f-entropy-1} and \ref{f-entropy-2}.

\begin{figure}[h!]
\includegraphics[scale=0.4]{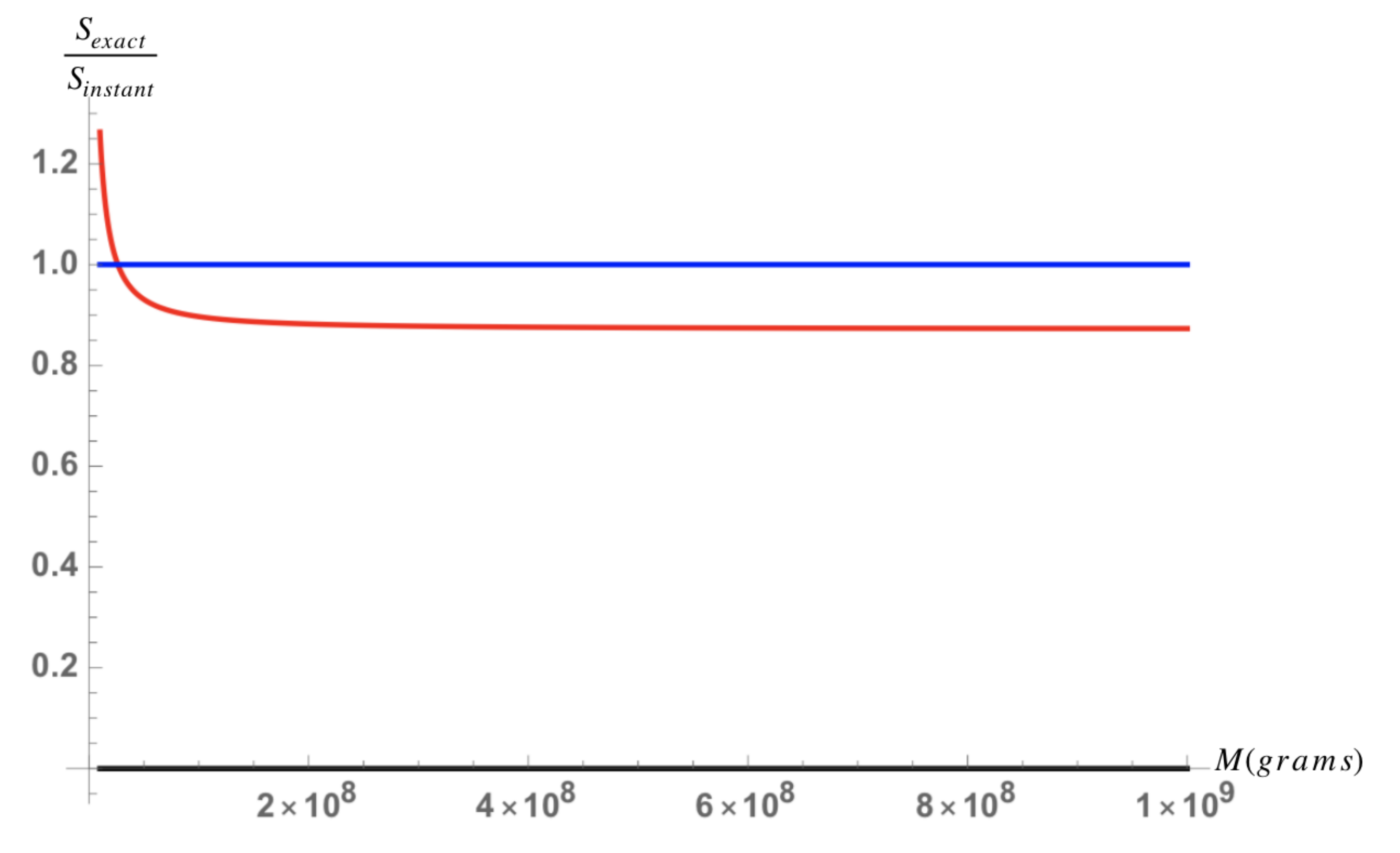}
\caption{The ratio of the entropy suppression factor of the exact fixed mass calculations (red) to the instant decay and change of 
the expansion regime approximation. The blue line describes  the hypothetical ratio equal to unity
}
\label{rat-del-inst-exct}
\end{figure}

The ratio of the entropy suppression factor of the exact fixed mass calculations 
{to that performed in the instant decay and change of 
the expansion regime approximation
as a function of mass for $\epsilon = 10^{-12}$ is presented in fig~\ref{rat-del-inst-exct}.
A rise of this ratio at small $M$ can be understood by underestimation of entropy release in the instant approximation. Indeed for $M$
smaller than the boundary value given by Eq.~(\ref{M-min-1}) the entropy release would be zero while the exact calculations  lead to 
nonzero result,  so their ratio would tend to infinity.}

\section{Extended mass spectrum  \label{s-extend}}

Let us now consider, instead of  delta-function, an extended  mass distribution:
\be 
\frac{dN_{BH}}{dM} = f(M,t) ,
\label{dN-dM}
\ee
where $N$ is the number density of PBH. Since PBHs are non-relativistic, their differential energy density is 
\be 
\frac{d\rho_{BH}}{dM} \equiv \sigma(M,t) = M f(M,t) ,
\label{rho-BH-2}
\ee

PBH created by the old conventional mechanism~\cite{ZN-PBH,Carr-Hawking} are supposed to have sharp, even
delta function mass spectrum. However, in several later works the mechanisms leading to extended mass spectrum 
have been worked out~\cite{AD-JS,DKK,INN,BLW}.

We assume that the number and energy densities of PBHs are effectively confined between $M_{min}$ and $M_{max}$.  
The value of $M_{max}$ should be effectively below the upper limit $M=10^9$ g, which is imposed by the condition that PBH
evaporation would not distort successful results of BBN-theory. However, a small fraction of PBHs may have masses higher 
than $10^9 $ g and  their impact on BBN can be interesting, though not yet explored in full.

The minimal value of PBH mass $M_{min} $ should be higher than $M_1^{min}$ given by eq. (\ref{M-min-1}) to ensure validity
of the assumption $\tau_{BH} \geq t_{eq}$ necessary for the entropy suppression fraction be larger than 1 else the impact of masses below $M_{min}$ would be inessential.


Let us parameterize the value of PBH mass using dimensionless parameter $x$ such that
$M_{BH}= x {M}_{0}$, where $M_0$ is  the average value of the mass density distribution or the value where
$\sigma (M,t)$ reaches maximum,  and
$x$ is a dimensionless number being non-zero in the limits: 
\be
x_{min} \equiv M_{min}/M_0 \le x \le x_{max} \equiv M_{max}/M_0 .
\label{x}
\ee

We define now the dimensionless "time"  $\eta$ as $\eta=t/ \tau(M_0)$
where $\tau ( M_0) \equiv \tau_0$ is the life time of PBH with mass $M_0$.
All the PBHs have different masses and hence their life-times (\ref{tau-BH})
and the moments of formation (\ref{t-in}) are different.

The evolution of the differential energy density of PBHs,  is governed  by the equation:
\be 
\dot \sigma (M, t)=-\left[ 3H+\Gamma(M)\right] \sigma (M,t),
\label{en-den}
\ee
where $\Gamma(M) = 1/\tau (M)  = m_{Pl}^4 /(C M^3)$, see eq. (\ref{tau-BH}).

In terms of dimensionless time $\eta$, the above expression takes the form:
\be 
\frac{d\sigma}{d\eta}  \equiv \sigma' =-\left[3H\tau_0+\left(\frac{M_0}{M}\right)^3\right] \sigma 
\label{en-den1}
\ee
The initial value of $\eta$ is the moment of BH formation. It depends upon $M$ and, 
according to  eq. (\ref {eta-in}),  is equal to
\be 
\eta_{form} (M) =\frac{m_{Pl}^2 M}{C{M_0}^3}
\label{eta-in-M}
\ee
Evidently $\sigma (M) = 0$ when $\eta (M) <\eta_{form}$. 

The equation describing evolution of the energy density of relativistic matter now takes the form:
\be
\frac{d \rho_{rel}}{d\eta} \equiv \rho_{rel}'  
= - 4H\tau_0  \rho_{rel}  + \int dM   (M_0/M)^3 \sigma (M) .
\label{rho-rel-eq-2}
\ee

In analogy with the previous section we introduce the red-shift function normalized to the value of the scale factor when
the least massive PBH was formed: 
\be
z (\eta) = a(\eta) / a\left[ \eta_{form}(M_{min}) \right]
\label{z-2}
\ee
The evolution of $z (\eta) $ is determined by the equation,
 analogous to Eq.~(\ref{dz-deta}): 
\be
\frac{dz}{d\eta} = H \tau_0 z
\label{dz-deta-1}
\ee
with the Hubble parameter now given by
\be 
\frac{3H^2  m_{Pl}^2}{8\pi} = \rho_{rel} + \rho_{BH} =
\rho_{rel}+ \int dM \sigma (M),
\label{hubble-of-M}
\ee

Eq.~(\ref{en-den1}) has the following solution 
\be
\sigma (M,\eta) = \theta \left(\eta - \eta_{f} (M) \right) \sigma (M,\eta_{f}) 
\exp \left[ -(\eta - \eta_{f} (M) ) 
\left(\frac{M_0}{M}\right)^3 \right] \left( \frac{z(\eta_{f} (M))}{ z(\eta)} \right)^3 ,
\label{sol-r}
\ee
where for brevity we have introduced the new notation $\eta_f \equiv \eta_{form}$,
 the theta-function ensures vanishing of the solution for $\eta < \eta_{f}$, 
and the initial value of the PBH  density at the moment  of formation
$\sigma (\eta_f (M))$ (\ref{eta-in-M}) 
is determined by the fraction $\epsilon (M)$
of the energy density of PBH with mass $M$ with respect 
to the energy density of the relativistic matter at the moment
of PBH formation:
\be
\sigma (M, \eta_f(M) )= \epsilon (M) \rho_{rel} (\eta_{f} (M) ) / M ,
\label{r-in-of-M}
\ee
where $\epsilon (M)$ depends upon the scenario of PBH formation and will be taken below according to some reasonable assumptions.
In any case we assume that $\epsilon(M)$ vanishes if $M<M_{min}$ and $M>M_{max}$.

We assume that in the time interval $\eta_f (M_{min}) <  \eta < \eta_f (M_{max})$ the total fraction of PBH mass density is negligibly 
small in comparison with the energy density of relativistic matter, and so the  expansion regime is the non-disturbed relativistic one, see 
eqs.~(\ref{rho-rel}, \ref{a-rel}). Accordingly using eq. (\ref{t-in}), we find that the energy density  of relativistic matter at the
moment of the creation of the "first" lightest black holes is 
\be
\rho_{rel} (t_{in}) = \frac{3}{32\pi} \,\frac{m_{Pl}^6}{M_{min}^2} .
\label{rho-rel-in}
\ee 
If the energy density of PBH remains small in comparison with that of relativistic matter till formation of the heaviest PBHs,
then the last term in the r.h.s. of eq,~(\ref{rho-rel-eq-2}) can be neglected and thus
in the time interval $\eta (M_{min}) < \eta < \eta (M_{max}) $  the energy density $\rho_{rel}$ evolves as
\be
\rho_{rel} =  \frac{3}{32\pi} \,\frac{m_{Pl}^6}{M_{min}^2}\frac{1}{z(\eta)^4} .
 \label{rho-rel-of-eta}
 \ee

Hence the differential PBH energy density evolves as 
\be
\sigma(M, \eta) = \frac{3 m_{Pl}^6}{32\pi M M^2_{min}}\,\frac{\epsilon (M)}{z(\eta_f(M)) }
\frac{\theta (\eta - \eta_f(M))}{z^3(\eta) \exp\left[ (M_0/ M)^3  (\eta - \eta_f(M) )  \right] } .
\label{sigma-of-eta}
\ee
In this  equation $\eta$ runs in the limits $\eta (M_{min}) < \eta < \eta (M_{max}) $ or 
$\eta_f (M)< \eta < \eta (M_{max})$, depending upon which lower limit is larger. 

Since $ (M_0 / M)^3 \eta_f (M) = m_{Pl}^2 / (C M^2) \ll 1$, for any $\eta$, we may expand the exponent as
\be
 \exp \left[ -(M_0/ M)^3  (\eta - \eta_f(M) )  \right]   =  \exp\left[ - (M_0/ M)^3  \eta   \right]  (1+  m_{Pl}^2 / (C M^2) )
 \label{exp-expand}
 \ee

Due to the necessity to integrate over $M$ the relevant evolutionary equations are integro-differential and the
numerical calculations generally become quite cumbersome. 
However, we can consider some simplified forms of the initial
mass distribution of the PBH for which the integrals over $M$ can be taken analytically and after that the 
differential equations can be quickly and simply solved. Using such toy models we can understand essential
features of the entropy  production by PBH with extended mass spectrum. Unfortunately we could not  find 
a workable toy model for a realistic log-normal mass spectrum, see ref. \cite{SMPBH}. Nevertheless the spectra which
allows for analytic integration can be quite close numerically to realistic log-normal one. 

We consider a couple of illustrative examples in what follows, assuming  that the function 
\be
F(x)=\epsilon (M)/z(\eta_f(M))
\label{F-def}
\ee
is confined between $x_{min} = (M_{min}/M_0)  $ 
and $x_{max} = (M_{max}/M_0)$. For simplicity we assume that  $F(x)$  is a 
polynomial function of integer powers of $x$, though the latter is  is not necessary.

We take two examples for $F$:
\be
F_1 (x) = \epsilon_0 /(x_{max} -x_{min})
\label{F1}
\ee
for $ x_{min} < x < x_{max}$ and $F_1 = 0$ for $x$ outside of this interval. Evidently $x=1$
should be inside this interval.

Another interesting form of $F$ is
\be
F_2  (x) = \frac{\epsilon_0}{N} \,a^2\,b^2  (1/a - 1/x)^2\,(1/x - 1/b)^2.
\label{f-2}
\ee
Here $N$ is the normalization factor, chosen such that the maximum value of $F_2 /\epsilon = 1$

This function vanishes at $x=x_{min} \equiv a$ and $x=x_{max} \equiv b$, with vanishing derivatives at these points, and 
$F_2 $ being identically zero  outside of this interval. 
$F_2$ reaches maximum at $x_0 = 2 a b /(a+b)$:
\be 
F_2^{(max)} = \frac{\epsilon_0}{16} N a^2 b^2 \left( \frac{1}{a} - \frac{1}{b} \right)^4 = 1.
\label{F2-max}
\ee

\begin{figure}[h!]
\includegraphics[scale=0.4]{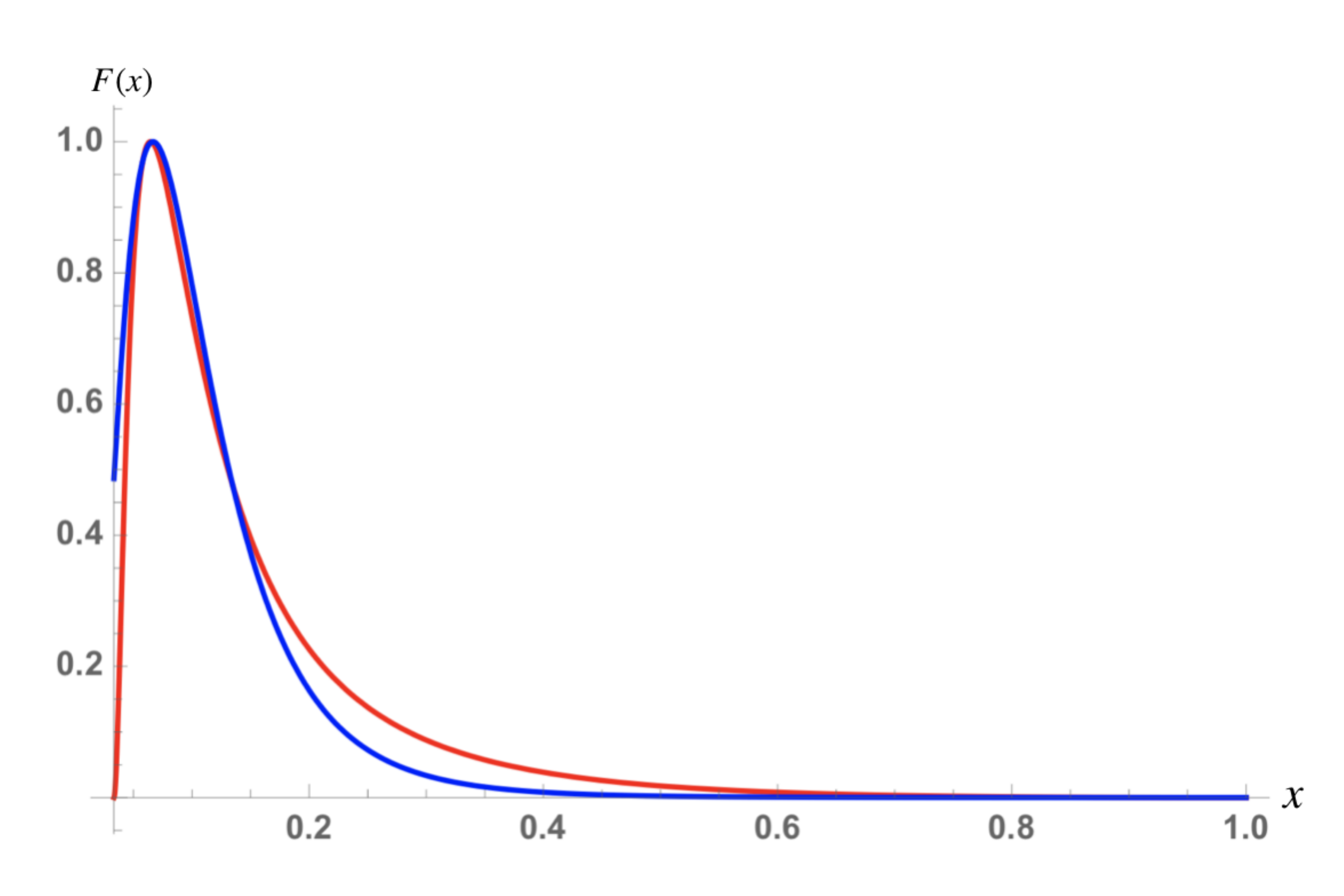}
\caption{The model mass spectrum function $F_2$ (red) and the log-normal spectrum (blue) as functions of $x = M/M_0 $.
}
\label{rat-F2-FLN}
\end{figure}

$F_2$ can  be quite close numerically  to the log-normal distribution with a proper choice of parameters. 
As a working example we take $a =1$, $b=30$ and compare $F_2$ with the log-normal function:
\be
F_{LN} =\epsilon  \exp[-1.5(\log^2 (15 x ))]
\label{F-LN}
\ee
 With the chosen parameters $F_2(x)$ and $F_{LN} (x) $ are presented 
 in Fig.~\ref{rat-F2-FLN}

There are two following integrals, which enter the evolution equation (\ref{hubble-of-M}) and (\ref{rho-rel-eq-2}):
\be
I_0 = \int dM \sigma (M,\eta)
\label{I-0}
\ee
and
\be
I_3 = \int dM \left(\frac{M_0}{M} \right)^3  \sigma (M,\eta).
\label{I-3}
\ee 
We can calculate them explicitly making some simplifying assumptions about the form of $F$ (\ref{F-def}),
which are discussed in the following subsections.

\subsection{Calculations for the flat spectrum \label{ss-flat}} 

Here we find the entropy suppression factor for the "flat" $F(x)$:
\be 
F_1 (x) = \frac{\epsilon (M)}{z(\eta_f(M))} = \frac{\epsilon_0}{b  - a} = const
\label{eps-flat}
\ee
for $a\equiv  x_{min} < x < b \equiv x_{max}$ and $F_1 (x) = 0$ outside this region.  
{Parameters $a$ and $b$ here and in what follows, Eq.(\ref{F2-x-LN}),
evidently define the width of the mass spectrum, so there is some but rather mild dependence
on them. Since there is no essential difference between the entropy suppression for extended and 
delta function mass spectra, the variation of $a$ and $b$ is not of much importance.
 }

Using eq.~(\ref{sigma-of-eta}) we find:
\be 
I_0^{(1)} &= & \int_{M{min}}^{M_{max}} dM \sigma (M,\eta) =
\frac{3 m_{Pl}^6 \epsilon_0}{32\pi z^3(\eta) M^2_{min}(b-a)} 
 \int \frac{dM}{M}
\frac{\theta [\eta - \eta_f(M)]}{ \exp\left[ (M_0/ M)^3  (\eta - \eta_f(M) )  \right] } = \nonumber \\
&=& \frac{ K(\eta)}{b-a} \int_{a}^{b} \frac{dx}{x} \frac{\theta [\eta - \eta_f(M)]}{ \exp\left[ x^3 (\eta - \eta_f(M) ) \right] }
\equiv \frac{K(\eta)}{b-a}\, j_{(10)}  (a, b, \eta,\eta_f) ,
\label{int-F1-0}
\ee
where $x=M_0/M$ and 
\be
K(\eta)  = \frac{3 m_{Pl}^6 \epsilon_0}{32\pi z^3(\eta) M^2_{min}} .
\label{K-of-eta}
\ee

\be 
I_3^{(1)} &=& \int_{M_{min}}^{M_{max}} dM \left(\frac{M_0}{M}\right)^3 \sigma (M,\eta) = 
\frac{K(\eta)}{b-a}\,\int_{x_{min}}^{x_{max}} \frac{dx}{x^4} 
\frac{\theta [\eta - \eta_f(M)]}{ \exp\left[  x^3  (\eta - \eta_f(M) )  \right] } \nonumber\\
&\equiv& \frac{K(\eta)}{b-a}\, j_{13} (x_{min},x_{max}, \eta, \eta_f) .
\label{int-F1-3}
\ee

We take integrals $j_{(10)}$ and $j_{(13)}$  analytically,
using Mathematica, and substitute them into equations 
(\ref{eta-in-M}) and (\ref{rho-rel-eq-2}), and (\ref{z-2}), which solve numerically. Since $\eta_f (M) \ll \eta$ in almost all integration interval
we neglect $\eta_f$, see also eq.~(\ref{exp-expand}). The results are presented in appendix B.

We will search for the solution as  it is done in sec.~\ref{s-extend} taking $\rho_{rel}$ in the form:
\be
\rho_{rel} =y_{rel} \, \rho_{rel}^{(in)}  / z^4 ,
\label{rho-rel-ex}
\ee
where $\rho_{rel}^{(in)}  = 3 m_{Pl}^6 /( 32 \pi M^2_{min})$ and so $y_{rel}$  and  $z$ satisfy the equations:
\be 
y' _{rel} &=& \epsilon_0\, z(\eta) j_{(13)}.
\label{dy-ex}\\
 z'(\eta) &=& \frac{C M_0^3}{2 m_{Pl}^2 M_{min}} \left( \frac{y_{rel}}{z^4} + \frac{\epsilon_0}{z^3} \, j_{(10)} \right)^{1/2}.
\label{dz-ex}
\ee

In analogy with eq.~(\ref{w}) we introduce new function $W_e$ according to
\be 
z = \sqrt{W_e}/\epsilon_0 .
\label{w-e}
\ee
and obtain:
\be 
\frac{d W_e}{d\eta} &=&  \frac{C \epsilon_0^2M_0^3}{m_{Pl}^2 M_{min}} \left( y_{rel} + {\sqrt{W_e}}\,j_{(10)}\right)^{1/2} \equiv
 \frac{C \epsilon_0^2M_0^2}{m_{Pl}^2 a} \left( y_{rel} + {\sqrt{W}_e}\,j_{(10)}\right)^{1/2}
\label{dWe} \\
\frac{dy_{rel}}{d\eta} &=& \sqrt{W_e}\, j_{(13)}
\label{dt--We}
\ee
with the initial conditions $W_{e}^{(in)} = \epsilon^2$ and $y_{rel}^{(in)} = 1$.

\begin{figure}[h!]
\includegraphics[scale=0.4]{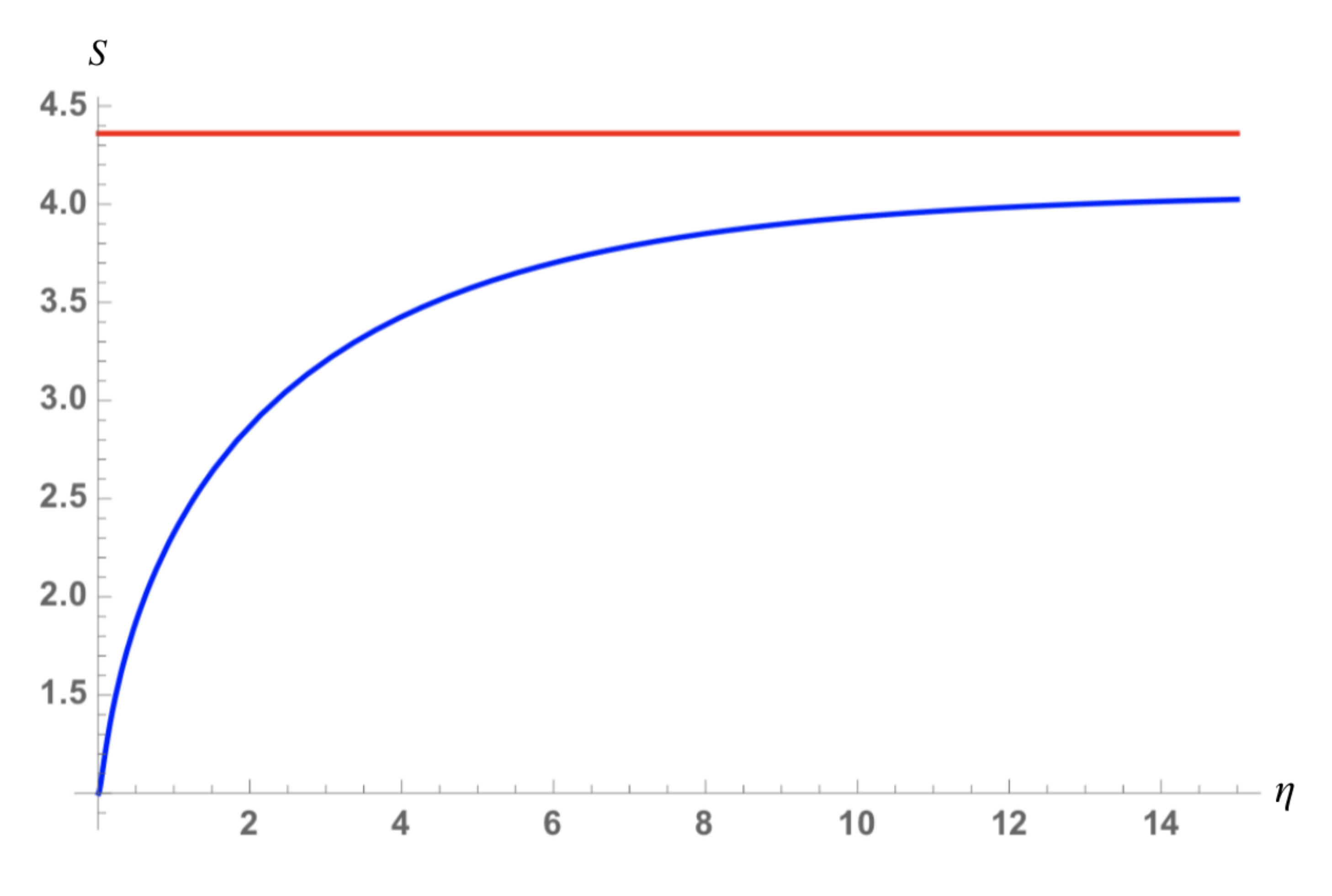}
\caption{The temporal evolution of entropy suppression $y_{rel}^{3/4}$ for flat mass spectrum (\ref{eps-flat}),
$M_{BH} = 10^7$ g and $\epsilon = 10^{-12}$ as a function of 
dimensionless time $\eta$ for $M_0 = 10^7 $ g,  $a= 1/3$, and $b = 4/3$ (blue). Red line is the entropy suppression factor approximately 
calculated in the instant approximation (\ref{suppress-2}).
}
\label{s-7-1/3}
\end{figure}

\begin{figure}[h!]
\includegraphics[scale=0.4]{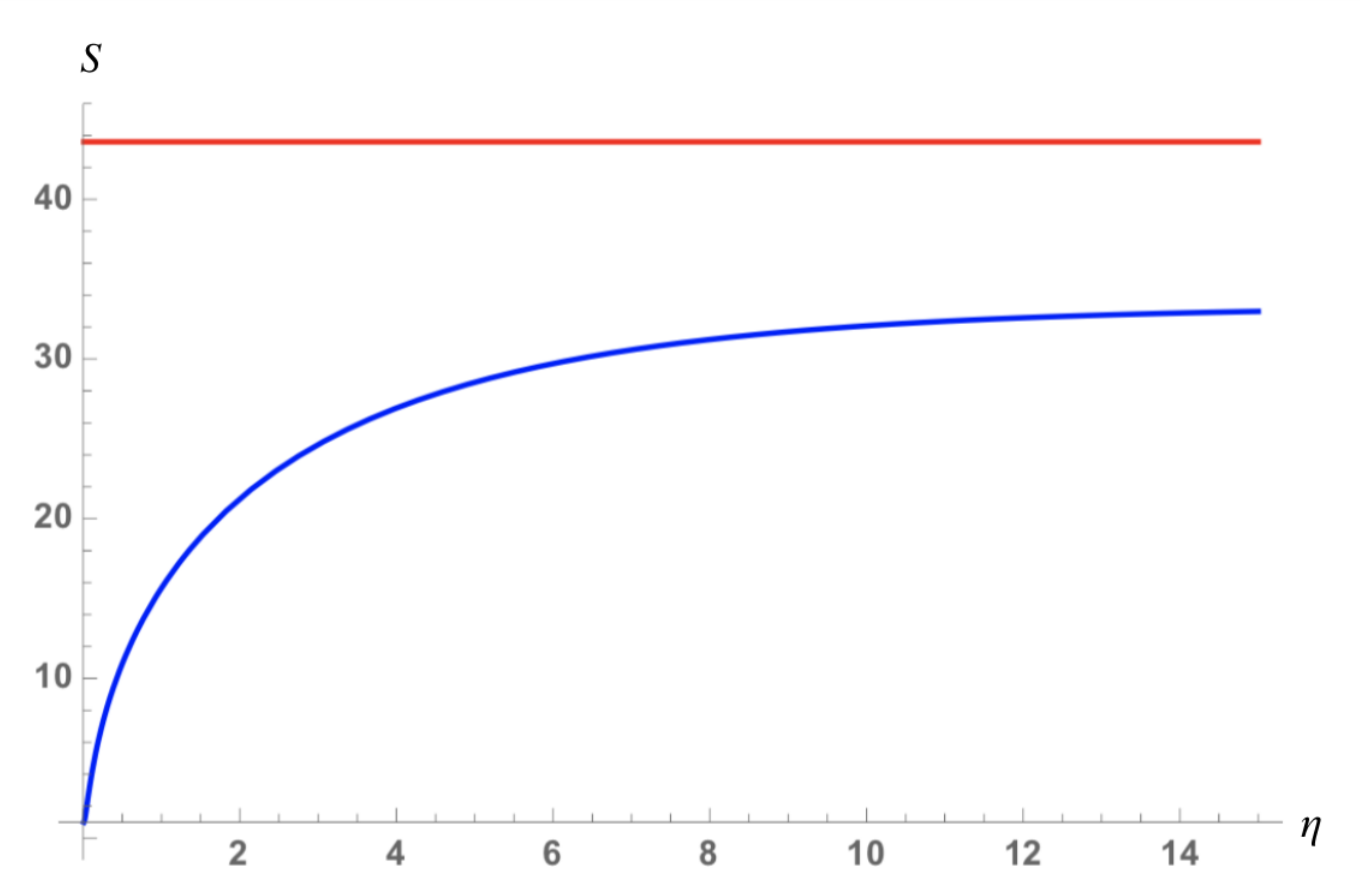}
\caption{The same as in fig.~\ref{s-7-1/3} but  with
$M_0 = 10^8 $ g} 
\label{s-8-1/3}
\end{figure}

\begin{figure}[h!]
\includegraphics[scale=0.4]{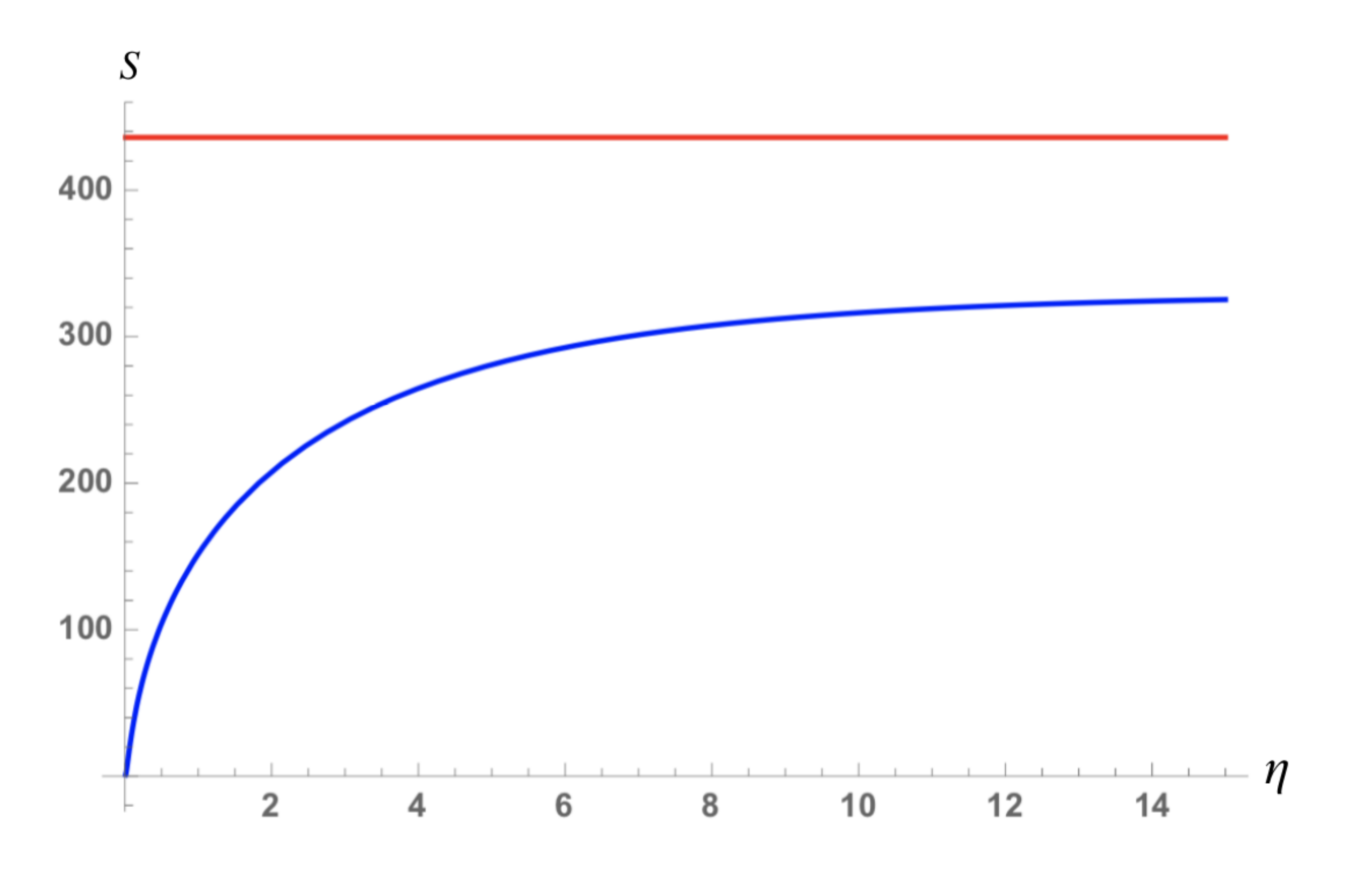}
\caption{The same as in fig.~\ref{s-7-1/3} but  with
$M_0 = 10^9 $ g} 
\label{s-9-1/3}
\end{figure}

\begin{figure}[h!]
\includegraphics[scale=0.4]{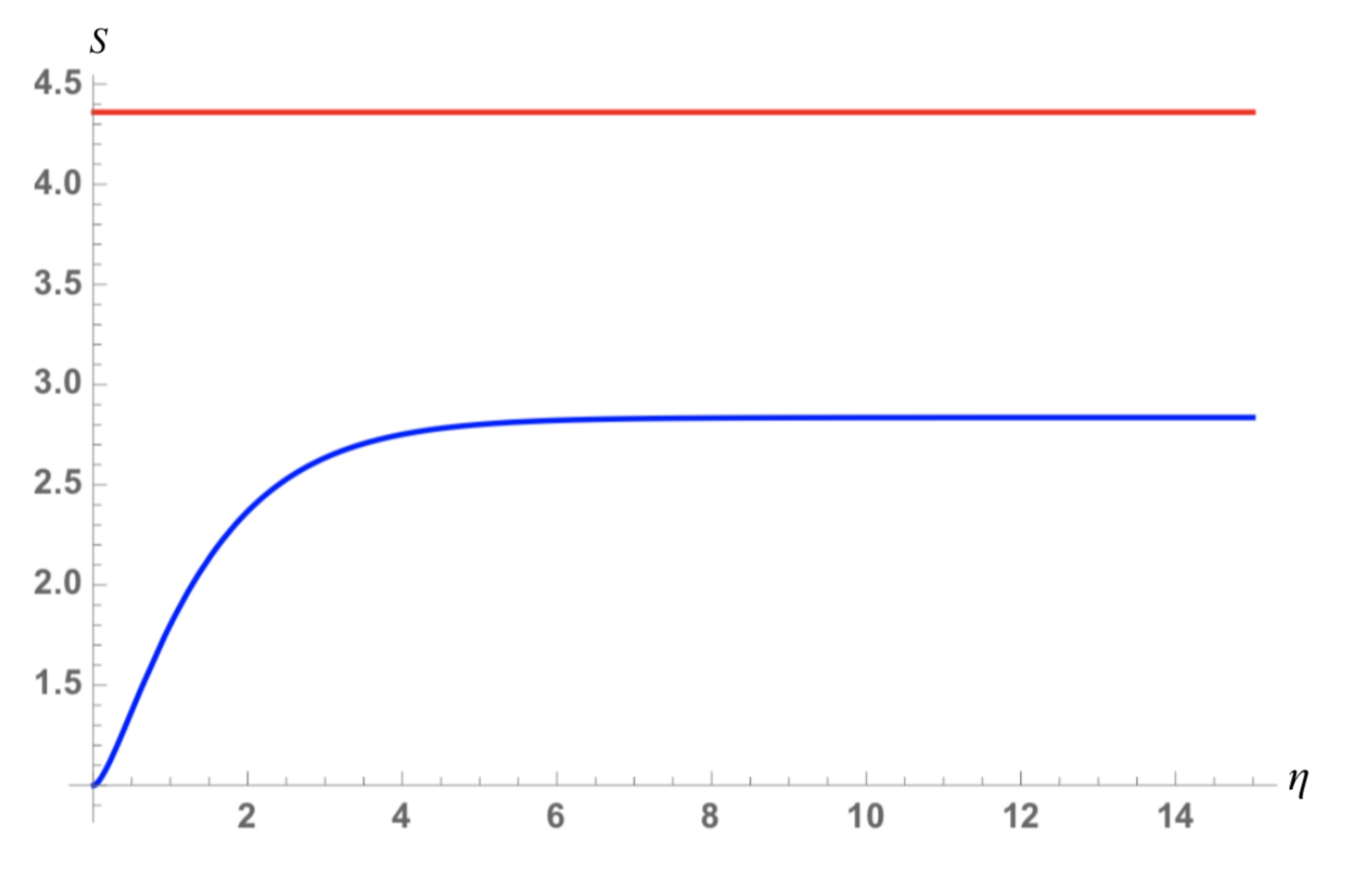}
\caption{The same as in fig.~\ref{s-7-1/3} but  with
 $a= 0.95 $, and $b = 1.05$
}
\label{s-7-95}
\end{figure}

\begin{figure}[h!]
\includegraphics[scale=0.4]{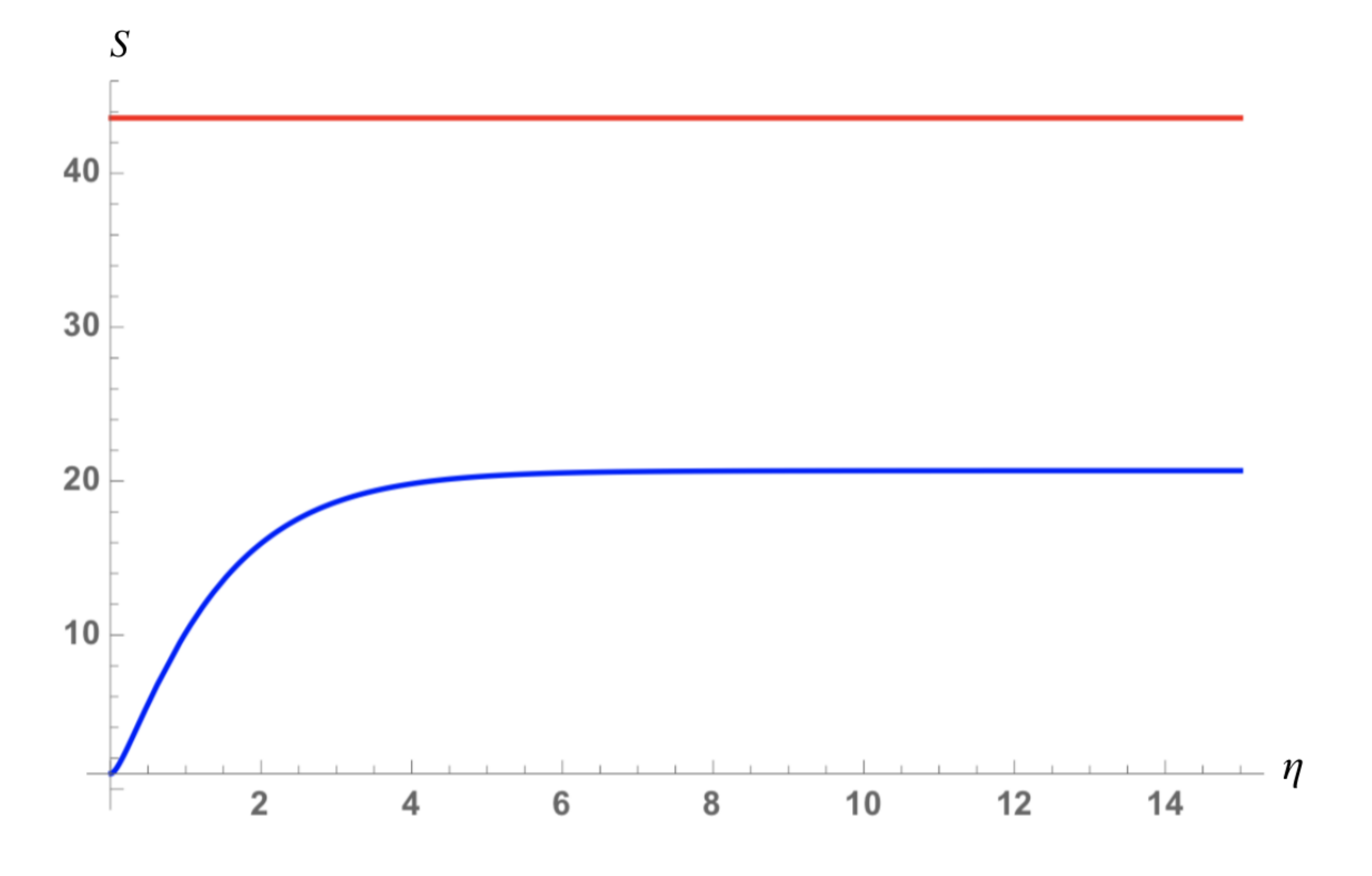}
\caption{The same as in fig.~\ref{s-7-1/3} but  with
 $a= 0.95 $,  $b = 1.05$, and $M_0 = 10^8$ g
}
\label{s-8-95}
\end{figure}

\begin{figure}[h!]
\includegraphics[scale=0.4]{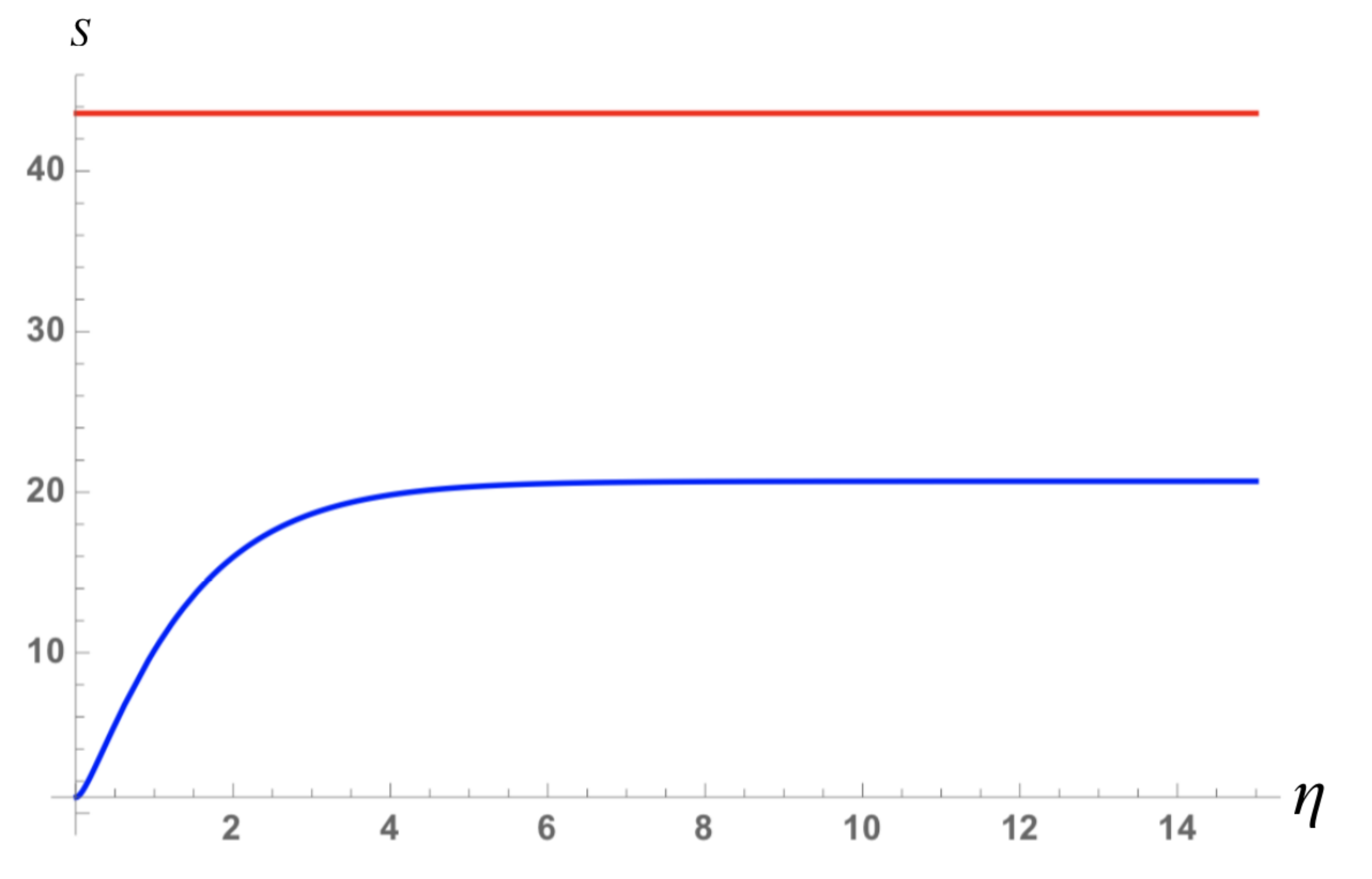}
\caption{The same as in fig.~\ref{s-7-1/3} but  with
 $a= 0.95 $,  $b = 1.05$, and $M_0 = 10^9$ g
}
\label{s-9-95}
\end{figure}

These equations can be integrated numerically. The asymptotic value of $y_{rel}^{3/4}$ at large $\eta$, which
is the entropy suppression factor according to eq.~(\ref{S})
is presented in figs.~\ref{s-7-1/3} - \ref{s-9-95}
all for $\epsilon =10^{-12}$ and $x_{min} = 1/3$ and $x_{max} = 5/3$.
The result is proportional to $M_{BH}$ and reasonably well agrees with  the approximate results calculated in
instant decay and instant change of regime approximations (\ref{suppress-2}).

\subsection{Calculations with almost log-normal  mass spectrum \label{ss-log-norm}}

Here we assume that
\be
F_2(x)=\epsilon (M)/z(\eta_f(M)) =  \frac{\epsilon_0 \,a^2\,b^2  (1/a - 1/x)^2\,(1/x - 1/b)^2}
{16  a^2 b^2 \left( {1}/{a} - 1/{b} \right)^4 }
\label{F2-x-LN}
\ee
Correspondingly equations (\ref{int-F1-0}) and (\ref{int-F1-3}) are modified by insertion of the factor $F_2 (x)$ into the integrands.
The expressions for $j_{(20)}$ and $j_{(23)}$ are presented in Appendix B.

\begin{figure}[h!]
\includegraphics[scale=0.4]{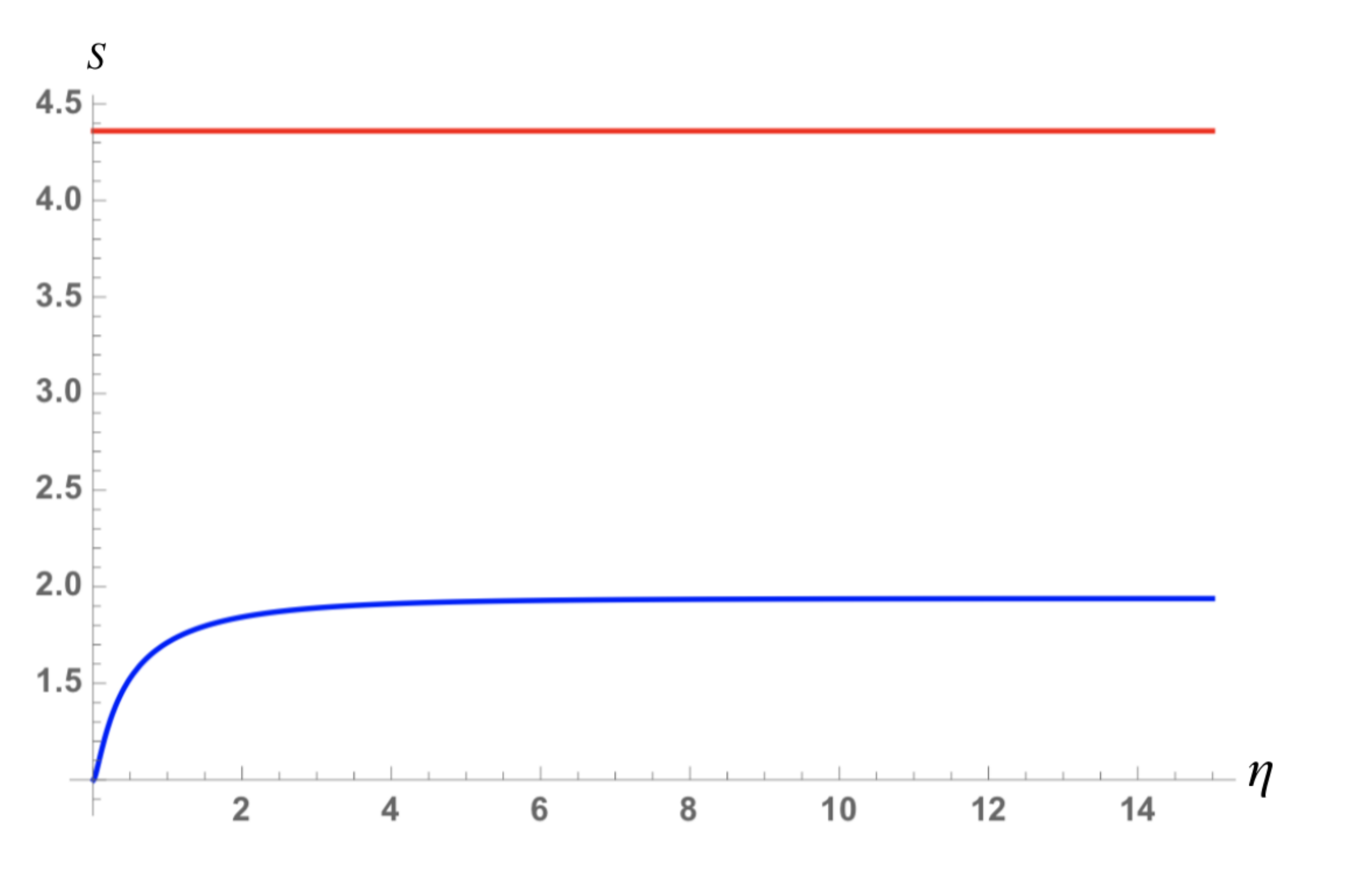}
\caption{The same as in fig.~\ref{s-7-1/3} but with the continuous  mass spectrum.
}
\label{s-9-95_}
\end{figure}

\begin{figure}[h!]
\includegraphics[scale=0.4]{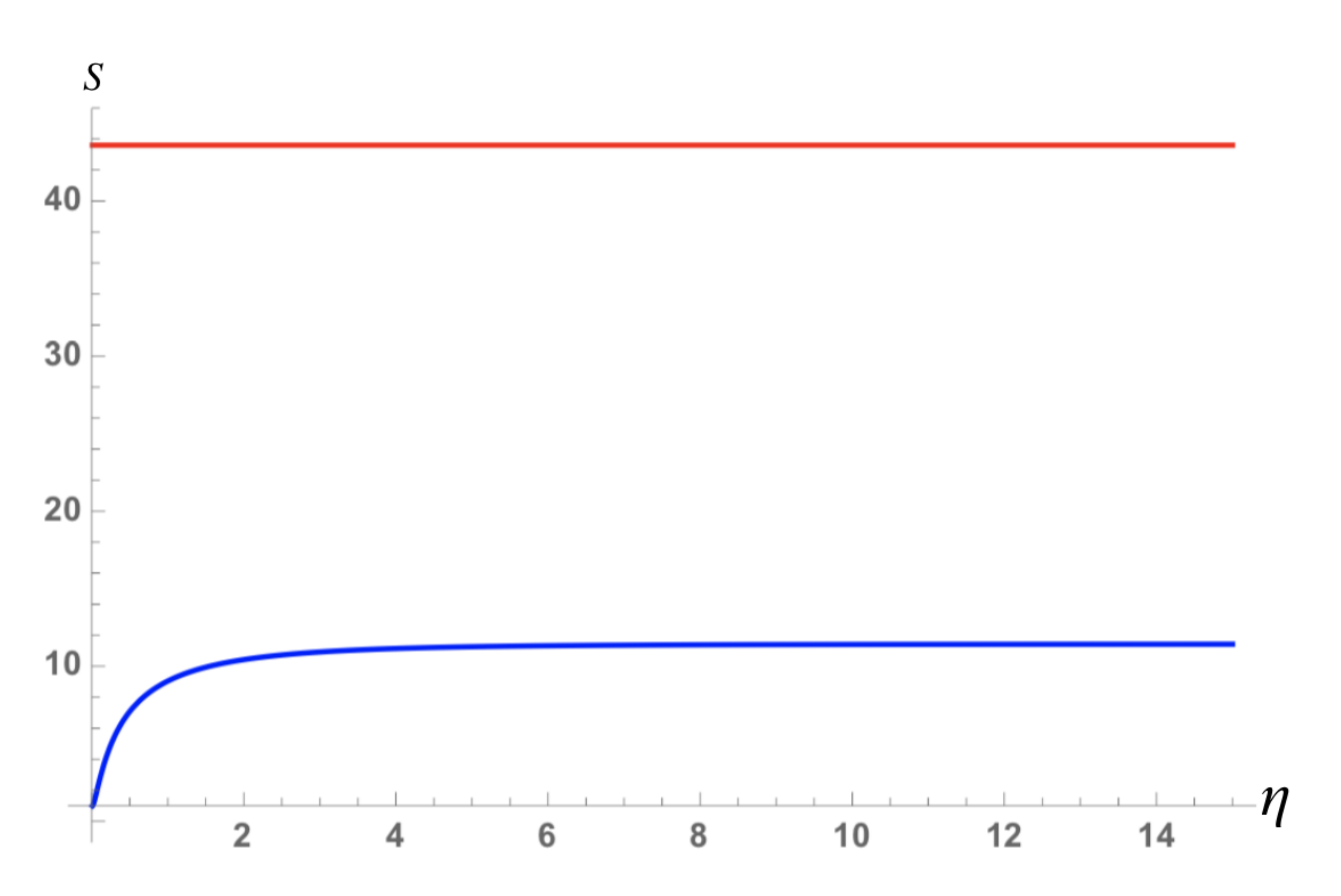}
\caption{The same as in fig.~\ref{s-7-1/3} but  with the continuous  mass spectrum.
and $M_0 =10^8$ g.
}
\label{s-9-951}
\end{figure}

\begin{figure}[h!]
\includegraphics[scale=0.4]{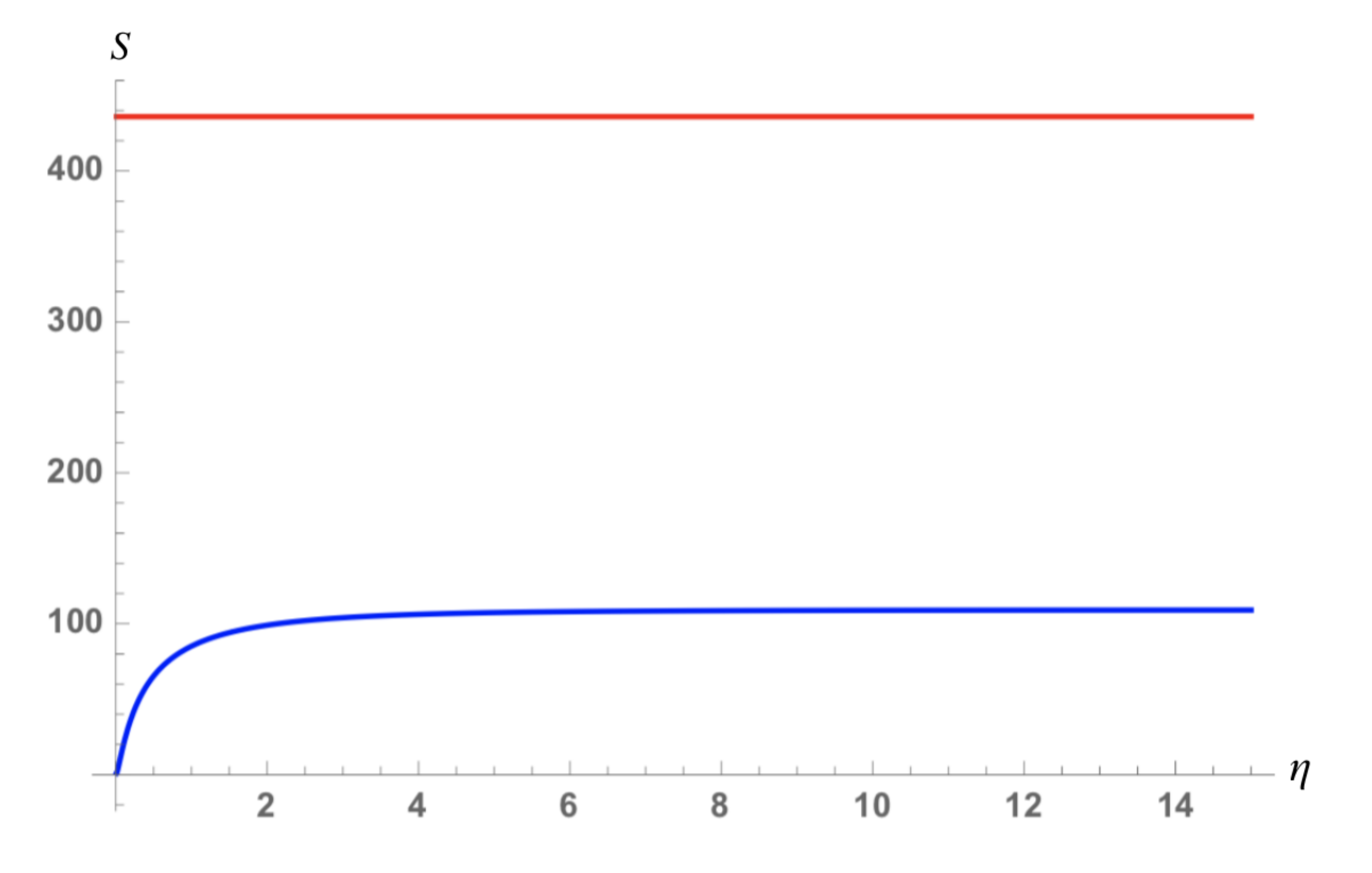}
\caption{The same as in fig.~\ref{s-7-1/3} but  with the continuous  mass spectrum
and $M_0 =10^9$ g.
}
\label{s-9-952}
\end{figure}

\begin{figure}[h!]
\includegraphics[scale=0.4]{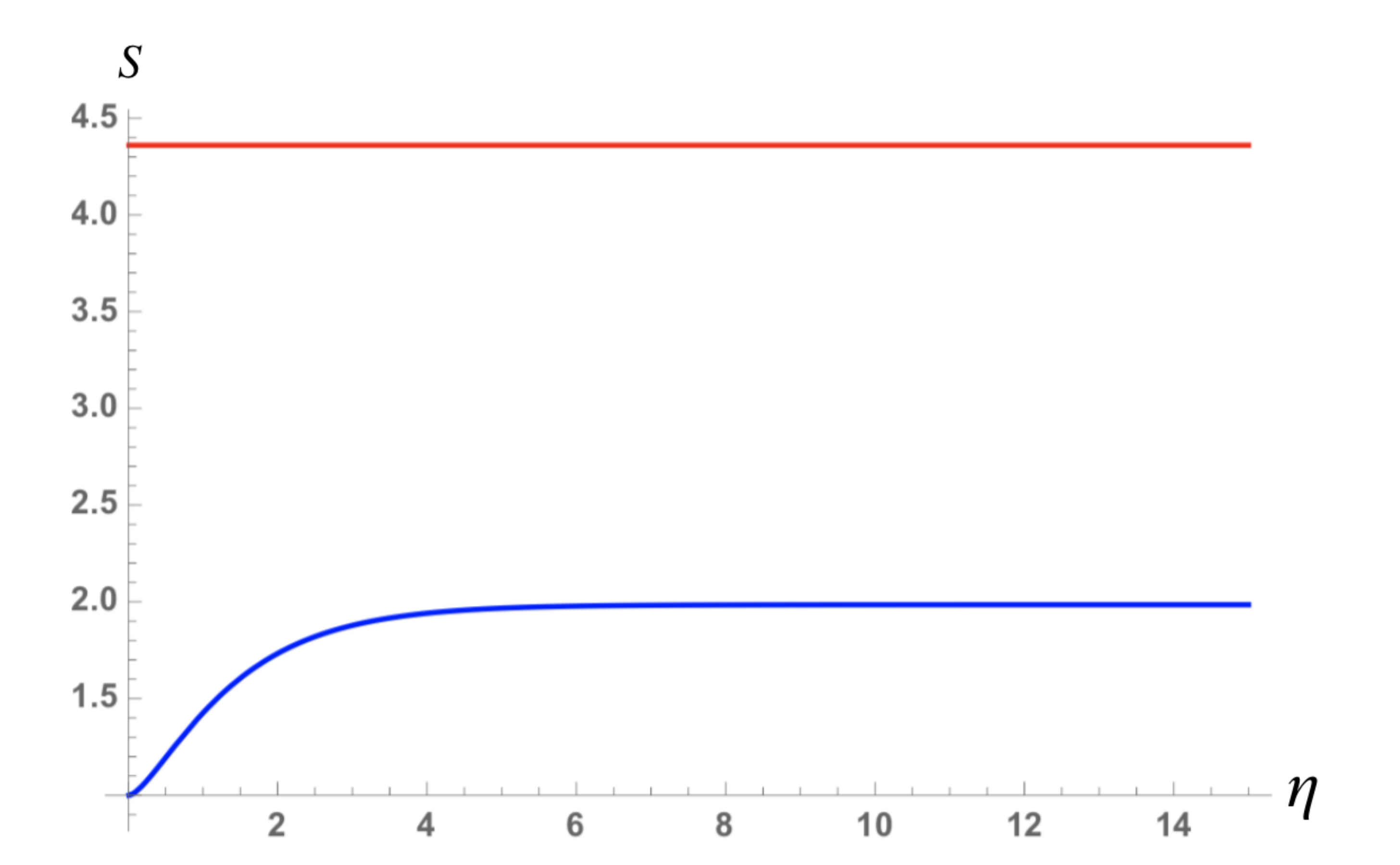}
\caption{The same as in fig.~\ref{s-7-1/3}  but  with the continuous  mass spectrum  and
$M_0 =10^7$ g,
 $a= 0.95 $,  $b = 1.05$, and $M_0 = 10^7$ g
}
\label{s-7-953}
\end{figure}

\begin{figure}[h!]
\includegraphics[scale=0.4]{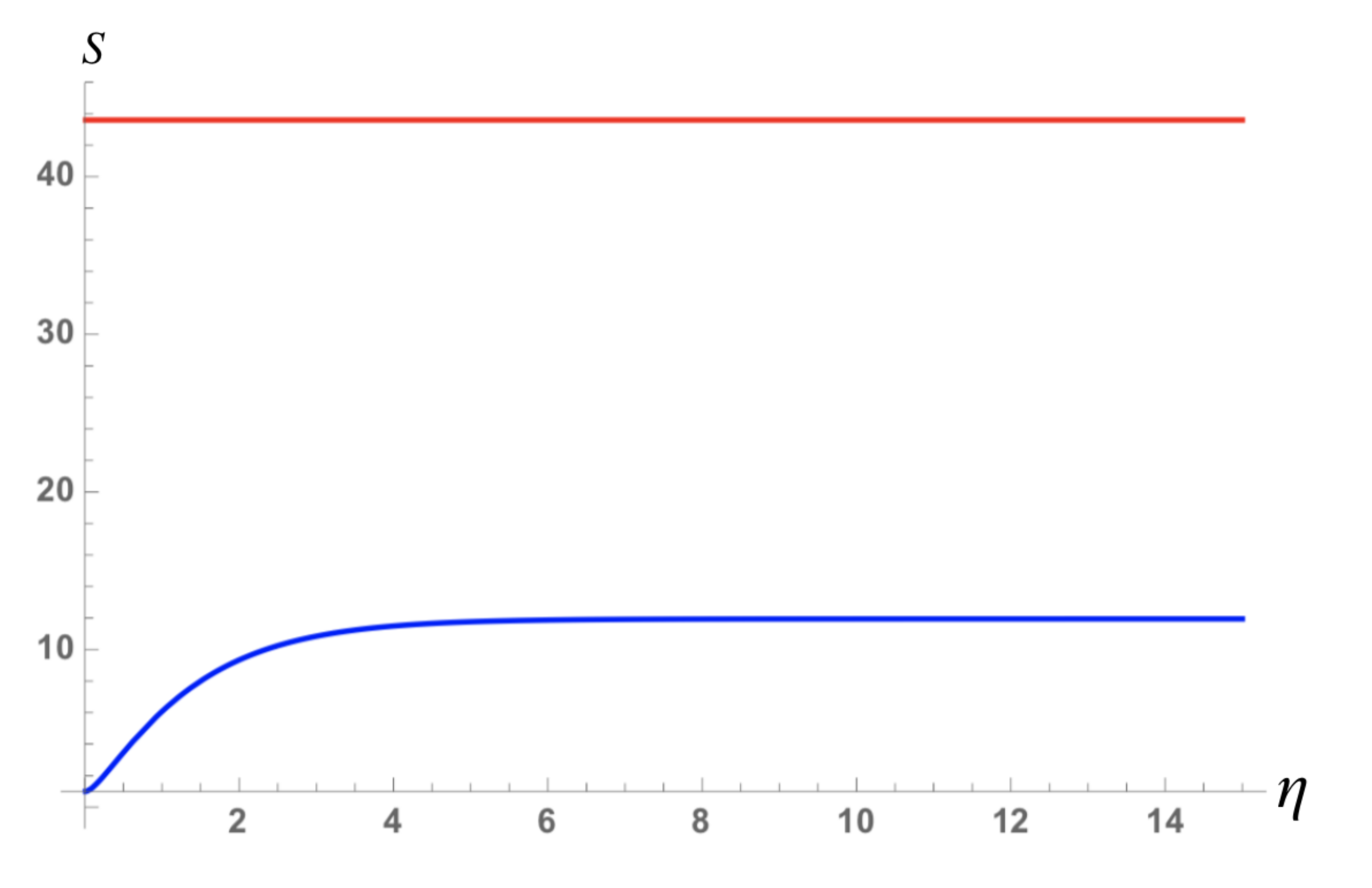}
\caption{The same as in fig.~\ref{s-8-1/3}   but  with the continuous  mass spectrum  and
$M_0 =10^8$ g,
 $a= 0.95 $,  $b = 1.05$.
}
\label{s-8-954}
\end{figure}

\begin{figure}[h!]
\includegraphics[scale=0.4]{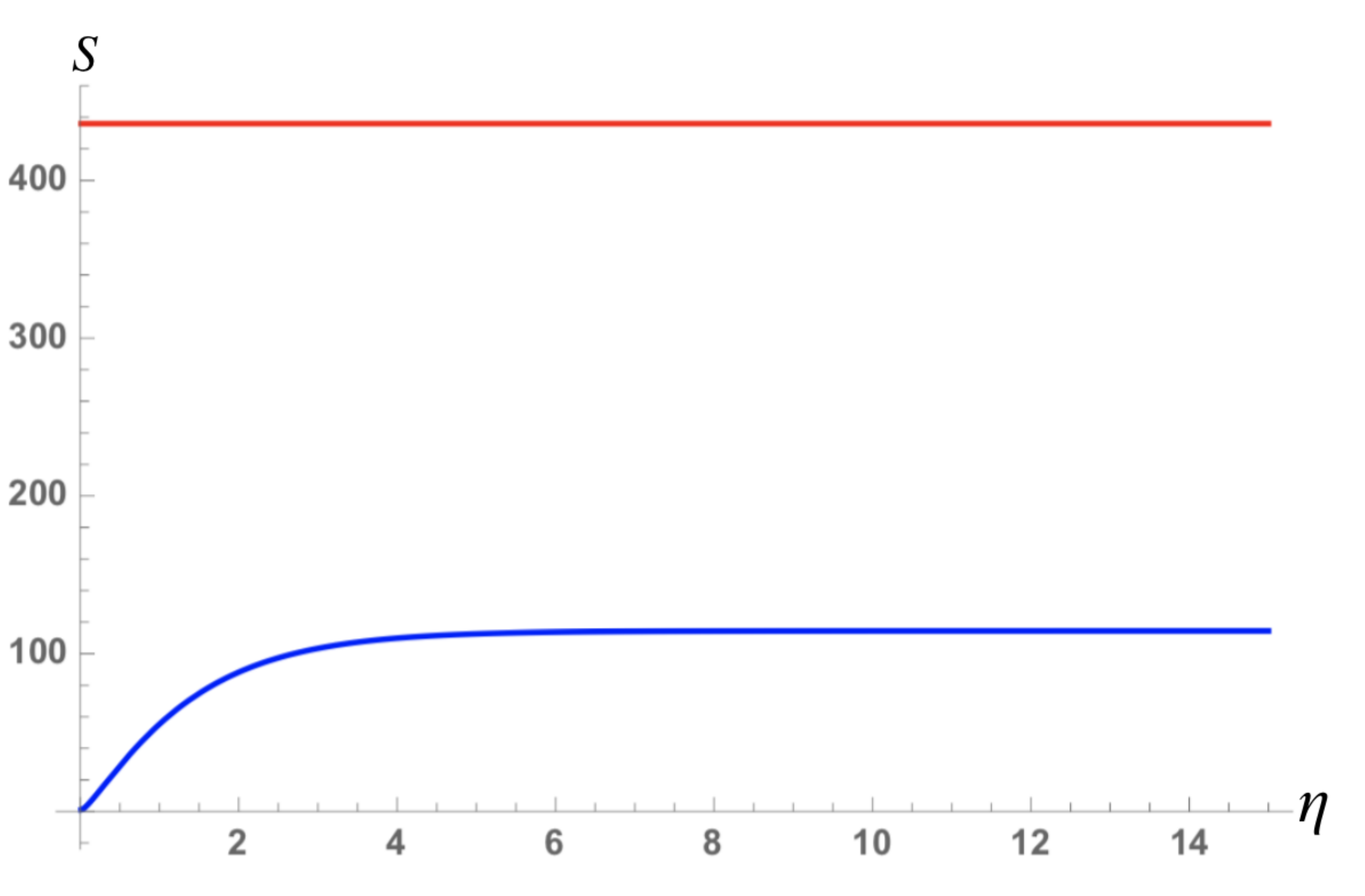}
\caption{The same as in fig.~\ref{s-8-1/3}   but  with the continuous  mass spectrum  and  
 $a= 0.95 $,  $b = 1.05$, and $M_0 = 10^9$ g
}
\label{s-8-955}
\end{figure}

\begin{figure}[h!]
\includegraphics[scale=0.4]{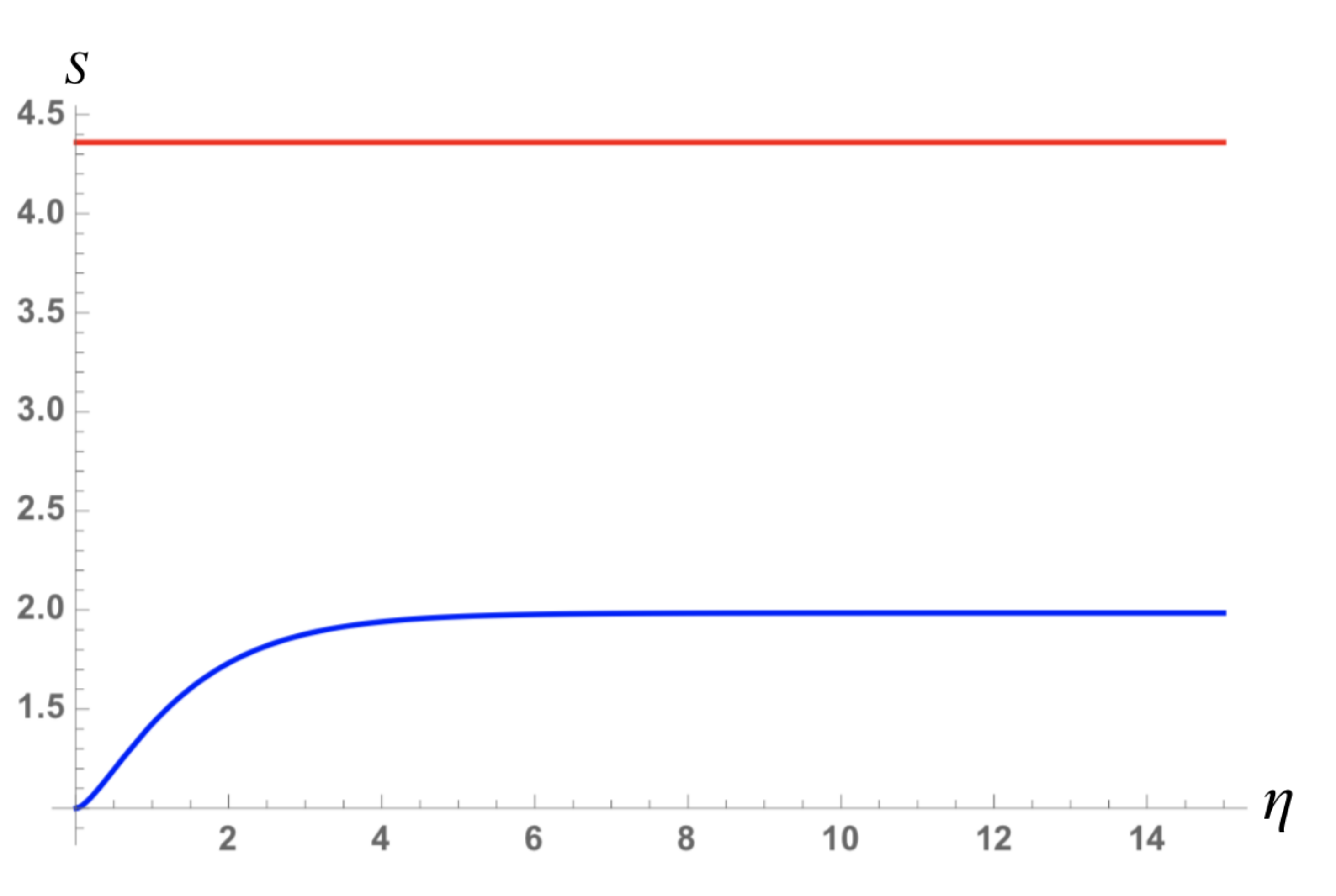}
\caption{The same as in fig.~\ref{s-8-1/3} but  with the continuous  mass spectrum  and  
 $a= 0.95 $,  $b = 1.05$, and $M_0 = 10^8$ g and $\epsilon = 10^{-13}$
}
\label{s-8-956}
\end{figure}

Evolution equations coincides with those in the previous subsection after the change $ j_{(10)} \rightarrow  j_{(20)} $ and
$ j_{(13)} \rightarrow  j_{(23)} $. The entropy suppression factor for the continuous mass spectrum and different  values of the 
parameters, indicated in the figure captions, are presented in figs. \ref{s-9-95_} - \ref{s-8-956}.

{We see that the entropy suppression factor for both studied here forms of
extended mass spectra, the rectangular
and more realistic log-normal one, behaves
as a function of the central value of the PBH mass and $\epsilon$
essentially similar to that calculated for the delta-function mass spectrum in Secs.~\ref{sec-instant}
and \ref{s-exact} and changes from the factor $ 2 - 3$ for $M=10^7$ g up to $100-300 $ for  $M=10^9$g.
 However, the 
comparison is ambiguous because it depends upon the normalization of the spectra, e,g, if we compare them at equal
mass densities of PBHs or at their equal number densities. It also depends upon the widths of the extended spectra.
Anyhow the outcome is the same by an order of magnitude. The dependence on $\epsilon$ is very accurately the same
as it was found in an analytical calculations of Sec.~\ref{sec-instant}.
}

\section{Conclusion \label{s-conc}}

As it is shown in this work, the suppression of thermal relic density or of  the cosmological baryon asymmetry may be significant  
if they were generated prior to PBH evaporation. In the simplified approximation of the delta-function mass spectrum of PBH,
instant decay of PBH, and instant change of the 
expansion regimes from the initial dominance of relativistic  matter to nonrelativistic BH dominance and back, the entropy suppression 
factor, $S$, can be calculated analytically, eq. (\ref{suppress-2}). Exact calculations but still with delta-function mass spectrum are in
very good agreement with the approximate one. 

The result is proportional to the product $ \epsilon M_{BH}$, and e.g. for $M_{BH} = 10^{9} $ g and $\epsilon = 10^{-12}$ the suppression 
factor is $S \approx 400$. 
The black hole mass equal to $10^9$ g is
the maximum allowed value of the early evaporated PBH mass permitted  by BBN , see conclusion below eq.~(\ref{T-heat}).
This statement is true if PBH dominated in the early
universe before the onset of BBN.  This could take place if  the minimal PBH mass is given by eq.~(\ref{M-min-1}).

The calculations with more realistic extended mass spectra of PBHs
show similar features of the suppression factor $S$, which is also proportional to $\epsilon$ and to
the central value  of the mass distribution. There is some dependence on the form of the spectrum and on the values of 
$M_{max}$ and $M_{min}$,  but they do not change our results essentially.

{The  significant restriction of the parameter space of the minimal supersymmetric model by LHC created some
doubts about dark matter made of LSP. Moreover, the usual  WIMPs with masses below teraeletron-volts seem to be 
excluded. The mechanism considered here allows to save relatively light WIMPs and open more options for SUSY dark matter. }

{
Similar dilution of cosmological baryon asymmetry by an excessive entropy release may look not so essential, because
theoretical estimates of the asymmetry is rather uncertain since they
strongly depends upon the unknown parameters of the theory at high energies. 
However, there are  a couple of exceptions for which the dilution may be of interest. \\
Firstly, there is the Affleck-Dine~\cite{AD-BG} scenario of baryogeneis, which naturally leads to the magnitude of the 
asymmetry, $\beta \sim 10^{-9}$ much higher than the observed one. The suppression by 1-2 orders of 
magnitude might be helpful, though not always sufficient.}

{Another example is baryo-thru-lepto genesis~\cite{BL-gen}, for a review see~\cite{BL-rev}. 
According to this model cosmological baryon asymmetry arise from initially generated lepton asymmetry, which is
generated by the decays of heavy Majoranna neutrinos. In some models the parameters of CP-violating decays 
of this heavy neutrino can be related to the CP-odd phases in light neutrino oscillations. Hence one can predict
the magnitude and sign of the lepton asymmetry. With the unknown dilution of the asymmetry the magnitude cannot 
be predicted but the sign probably can.}


\section{Acknowledgements}
Our work was supported by the RSF Grant 19-42-02004.

\newpage

\section{Appendix A}
 
We estimate here the density of stable supersymmertric relics produced in PBH evaporation and show that their
contribution to the cosmological dark matter is insignificant, due to very low density of the PBHs. To this end we
will present here a few simple estimates and numerical values.

The moment of PBH production with mass M is (\ref{t-in}):
\be
t_{in} = \frac{M}{m_{Pl}^2} = 2.5 \cdot 10^{-31}\, M_8 \,{\rm sec},
\label{t-in-sec}
\ee
 where $M_8 =M/(10^8 {\rm g})$.
 
By assumption at the moment of production PBHs make a small fraction $\epsilon \ll 1$ of the energy density of relativistic matter.
So the energy and number densities of PBH at $t = t_{in}$ are respectively:
\be 
\rho_{BH}^{(in)}=  \frac{3\epsilon}{32 \pi}\,\frac{m_{Pl}^6}{M^2},\,\, \,\,
n_{BH}^{(in)} =  \frac{3\epsilon}{32 \pi}\,\frac{m_{Pl}^6}{M^3}.
\label{rho-bh-in}
 \ee
 The energy density of the relativistic matter at $ t = t_{in}$ is:
\be 
\rho_{rel}^{(in)}=  \frac{3}{32 \pi}\,\frac{m_{Pl}^6}{M^2} =  \frac{\pi^2 g_*^{(in)}}{30}\,T^4_{in},
\label{rho-bh-in}
 \ee
 where $g_*^{(in)} \approx 100 $ is the number of relativistic species at $T=T_{in}$. 
  Correspondingly the temperature of the relativistic cosmological plasma
 at the moment of PBH production is equal to
 \be
 T_{in} \approx 1.72\cdot 10^{12}\,{{\rm GeV}}/{\sqrt{M_8}}.  
  \label{T-in}
 \ee

  The ratio on PBH number density to that of relativistic particles at the moment of creation can be estimated as:
 \be 
 r_{in} = \frac{n_{BH}^{(in)}}{n_{rel}^{(in)}} =  \frac{\rho_{BH}^{(in)}}{\rho_{rel}^{(in)}}  \frac{T _{in}}{0.3M} 
 =0.9\cdot 10^{-31}\epsilon_{12} M_8^{-3/2},  
  \label{r-in}
 \ee
 where $\epsilon_{12} = 10^{12} \,\epsilon$ and  $n_{rel} \approx 0.3\rho_{rel} /T$.
 
This ratio remains approximately constant till the PBH decay because both densities are
almost conserved in the comoving
volume up to the entropy release created by massive particle annihilation. As we see in what follows, the temperature 
of the relativistic matter at the moment of PBH decay is about 20-30 MeV and so at that moment $g_* \sim 10$. 
Hence the ratio $r$ drops down by factor 10. 
 
 The average distance between PBHs at the moment of their creation is
 \be  
 d_{in}^{(BH)} = \left(n_{BH}^{(in)}\right)^{-1/3} = 2.4\cdot 10^{-16}\,  M_8 \epsilon^{-1/3}_{12} \,{\rm cm} .
  \label{d-in}
 \ee
  At the moment of equilibrium, when densities of BH and relativistic matter became equal, the 
 average distance of BH separation was
 \be 
  d_{eq}^{(BH)} = d_{in}^{(BH)}/\epsilon = 2.4\cdot 10^{-4}\,  M_8 \,  \epsilon^{-4/3}\,{\rm cm}.
  \label{d-eq}
  \ee
 The temperature of the relativistic matter at the equilibrium  moment was
 \be
 T_{eq} = \epsilon T_{in} S_{eq}^{1/3} =  
   3.7 \epsilon_{12} \,M_8^{-1/2} \, {\rm GeV} ,
  \label{T-eq}
  \ee
  where $S_{eq}$ is the ratio of the number of particle species at $T= T_{in}$ to that at $T_{eq} \approx 10$:
   $ S_{eq} = g_* (10^5 {\rm GeV} )/g_* (3 {\rm GeV} ) =10 $.
    
 Since before the equilibrium the universe expanded in relativistic regime, when the scale factor rose as 
$ a(t) \sim t^{1/2} $,  the equilibrium is reached at the moment of time:
  \be 
  t_{eq} = t_{in}/\epsilon^2 = 2.5\cdot 10^{-7} M_8 \epsilon_{12}^{-2}\, {\rm sec} 
  \label{t-eq}
  \ee
  
  After that and till the moment of BH decay at 
  \be
  t=\tau = 
  30\,{M_{BH}^3}/{m_{Pl}^4}= 1.6\cdot 10^{-4} M_8^3 \,{\rm sec}
 \label{tau-BH-new}
\ee
 the universe expanded in matter dominated regime, $a(t) \sim t^{2/3}$. 
 So during this MD stage the scale factor rose as:
 \be
 z(\tau) \equiv \left(\frac{\tau}{t_{eq}} \right)^{2/3} = 
 74\,(\epsilon_{12}\cdot M_8 )^{4/3} . 
 \label{z-tau}
 \ee
  Correspondingly the energy density of PBHs just before the moment of their decay would be larger than
 the energy density of the relativistic background by this redshift factor, $z(\tau)$:
 \be
\frac{ \rho_{BH} (\tau)}{\rho_{rel}(\tau)} =  74\,(\epsilon_{12}\cdot M_8 )^{4/3} .
 \label{rho-BH-to-rho-rel}
 \ee
 The temperature of the relativistic background just before the BH  decay was
 \be
 T_{cool}  \equiv T _{rel} (\tau)=  T_{eq} /z (\tau) =
 50\,  \epsilon_{12}^{-1/3} M_8^{-11/6} \,{\rm MeV} .
 \label{T-of-tau}
 \ee 
 The temperature of the particles produced in the BH  decay is equal to:
\be 
T_{BH} = \frac{m_{Pl}^2}{8\pi M} = 10^5 M_8^{-1} \,{\rm GeV} 
\label{bar-E}
\ee
So the lightest supersymmetric particles (LSP) of the minimal SUSY model with the mass $m_X \sim 10^3$ GeV
should be abundantly produced  in the process of   the PBH evaporation with $T_{BH} \gg m_X$, 
contributing about 0.01-0.1  to  the total number of the produced particles.

 The average distance between PBH just before their decay was:
 \be
 d^{BH} (\tau) = d_{eq}^{(BH)} \cdot z(\tau) \approx 1.75 \cdot 10^{-2}\, M_8^{7/3}\,{\rm cm} .
 \label{d-of-tau}
 \ee
  The total number of energetic particles produced by the decay of a single BH is:
 \be
 N_{hot} \approx \frac{M_{BH}}{3 T_{BH}} = \frac{8\pi}{3}\,\left( \frac{M}{m_{Pl} } \right)^2 = 1.8\cdot 10^{26} M_8^2 .
 \label{N-hoit}
 \ee
 
 
 We assume the following model: as a result of BH instant evaporation each black hole turns into a cloud of 
 energetic particles with temperature $T_{BH} = 10^5 M_8^{-1}$ GeV,  with  
 radius $\tau_{BH}$, see e.g. eq,  (\ref{tau-BH-new}):
 \be 
  \tau_{BH} = 4.8 \cdot 10^6 M_8^3 \,{\rm cm}.  
  \label{tau-BH-cm}
  \ee 
  This radius is much larger than the average distance between the BHs (\ref{d-of-tau}) and the number 
  of PBHs in this common cloud is  
\be
N_{cloud}  = \left(\tau_{BH}/d_{BH}(\tau)\right)^3 = 2 \cdot 10^{25} M_8^7
\ee
 and their number density just before the decay was 
 \be
 n_{BH} (\tau) = d(\tau)^{-3} = 1.9\cdot 10^{5}  M_8^{-7}\,{\rm cm}^{-3} .
 \label{n-BH-of-tau}
 \ee 
 The density of hot particles with temperature $T_{BH}$, created by the evaporation of this set of black 
 holes is:
 \be
 n_{hot} = n_{BH} (\tau) \cdot N_{hot} = 3.4\cdot 10^{31}  M_8^{-5}\,{\rm cm}^{-3} .
  \label{n-part}
 \ee
 The density of cool background  particles with temperature $T_{cool}$ (\ref{T-of-tau}) is
  \be 
 n_{cool} = 0.1g_* T_{cool}^3 = 1.6\cdot 10^{37} \epsilon_{12}^{-1} M_8^{-11/2} {\rm cm^{-3}},
  \label{n-cool}
 \ee
 where we took $g_* =10$ at $T< 100$ MeV. Note that $n_{cool} \gg n_{hot}$.

The particles produced by PBH evaporation consist predominantly of some light or quickly decaying 
species and a little of stable lightest supersymmetric particles (or any other stable particles, would-be 
dark matter), 
  denote them as $X$. {Since by assumption $T_{BH}$ is higher than the SUSY mass scale, 
  the total number of all supersymmetric partners created through evaporation should be equal to the
 number of all other particles. Each SUSY partner produces one LSP ($X$-particle) in the process of its 
 decay and a few other particle species.  So the number of $X$s became roughly about 
 one per cent of the number of other particle number. More precise value is not of much importance here.
 This ratio further significantly dropped down 
 in the process of thermalization, see bellow
 }

 The ejected energetic particles propagate in the background of much colder plasma and cool down 
simultaneously heating the background. The cooling proceeds, in particular,  through the Coulomb-like scattering, 
so the momentum of hot particles decreases according to the equation
(the term related to the universe expansion is neglected there because the characteristic time scale of cooling
is much shorter than the Hubble time at $T\sim 100$ MeV):
\be
\dot E_{hot} = -\sigma v n_{cool} \delta E, 
\label{dot-p}
\ee  
where $\delta E$ is the momentum transfer from hot particles to the cold ones. The scattering cross-section can be approximated as 
$ \sigma = \alpha^2 g_* /| p_1 - p_2|^2 $. For massless particles
\be
q^2 \equiv (p_1 - p_2)^2 = -2 (E_1 E_2 - \vec p_1 \cdot \vec p_2 ) .
 \label{p1-p2}
\ee 
 Here $E_1$ and $E_2$ are the initial and final energies of cold particles, $E_1 \sim T_{cool}$ and
 $\delta E \equiv (E_2 - E_1) \sim E_2$. 
 For noticeable energy transfer large angle scattering is necessary, so $q^2  \sim E_1 E_2 $. Finally
 \be
 \dot E = 0.1 g_* T_{cool}^3 \alpha^2  /E_1 \approx 10^{-4} T^2_{cool} = 6\cdot 10^{18} {\rm MeV/sec}.
 \label{dot-E2}
 \ee
 Correspondingly the energy loss of hot particles of the order of  their temperature (\ref{bar-E}) would be
 achieved during very short time: 
 \be 
 t_{coot} \approx 10^{-10}\, {\rm sec}.
 \label{t-cool}
 \ee
 {Such a quick cooling is ensured by a huge number density of cool particles: there are about a million of cool particles
 over each hot one, see eqs. (\ref{n-part},  \ref{n-cool}) .
 }
 
 As a result of mixing and thermalization of two components, hot and cool, the temperature of the resulting plasma 
 would become:
 \be
T_{fin}  = T_{cool} \left( \rho_{hot}/\rho_{cool} \right)^{1/4} \approx 147 M_8^{-3/2}\,{\rm MeV}.
 \label{T-fin}
 \ee
 Correspondingly the total number density of relativistic particles {would be equal to:}
 \be
 n_{rel} = 0.1 g_* T_{fin}^3 = 4\cdot 10^{38}  M_8^{-9/2}  /{\rm cm}^3 .
 \label{n-rel-2}
 \ee
 
 { According to Eq.~(\ref{n-part}) the number density of $X$-particles immediately after evaporation
 should be about $ 10^{30}  M_8^{-5}\,{\rm cm}^{-3} $. After fast thermalization the ratio of number
 densities of $X$s to that of all relativistic particles becomes:
 \be
 n_{X} / n_{rel} = 3\cdot 10^{-9} .
 \label{n-X-to-n-tot}
 \ee
  }
 
 The evolution of the number density of X-particles is governed by the equation:
 \be{
  \dot n_X + 3H n_X = - \sigma_X^{(ann)} v n_X^2,}
 \label{dot-n_X}
 \ee
 where the inverse annihilation term is neglected because  hot particles from the 
 PBH evaporation cool down very quickly with characteristic time (\ref{t-cool}) and hence the plasma 
 temperature became much smaller than $M_X$. 
 {Evidently since $m_X \gg T_{fin}$ (\ref{T-fin}), the distribution of $X$-particles would be very much different
 from the equilibrium Bose-Einstein or Fermi-Dirac distributions but the kinetic equilibrium should be quickly established leading 
 to the distribution over energy close
 to the equilibrium ones with non-zero and {\it equal} chemical potentials of $X$ and anti-$X$, assuming zero charge
 $X/\bar X$ - asymmetry. If total kinetic and chemical equilibrium would be established, the number densities of $X$ 
 (and $\bar X$) would be extremely small and the problem of their over-abundance would not appear.  
 The key point here is the fast cooling of the plasma of the produced hot particles, much faster than the 
 cosmological expansion rate, see eq.~(\ref{t-cool})
 }
 
 {The Hubble parameter $H$ which enters  Eq.~(\ref{dot-n_X})  is given  by the expression:
 \be
 H= \left(\frac{8 \pi^3 g_*}{90}\right)^{1/2}\,\frac{T^2}{m_{Pl}} \approx  \frac{0.4\,T_{in}^2}{z^2 m_{Pl}},
 \label{H-2}
 \ee
 where $z= a_{in}/a$ is the ratio of the initial scale factor to the running one and for $T_{in}$ we
 take $T_{fin} $ given by Eq.~(\ref{T-fin}). Hopefully it will not lead to confusion.
 }
 
 {
 Introducing $r = n_Xz^3$ and changing the time variable to $z$, we arrive to the equation:
\be
\frac{dr}{dz} = -  \sigma_{ann} v\,\frac{r^2}{H z^4} = 
-\frac{\sigma_{ann} v m_{Pl}} {0.4 T^2_{in}} \,\frac{r^2}{z^2},
\label{dr-dz}
\ee
 which is easily solved leading to
 \be
 n_X = \frac{n_{in}}{z^3  \left( 1 - 1/z \right)} \rightarrow \frac{1}{Q z^3},
\label{nX-of-z}
\ee
 where $Q= (\sigma v\, m_{Pl}) /(0.4 T_{in}^2)$.
 }

 { The total annihilation cross-section can be fixed by the condition that $X$-particles are
the dominant carriers of the cosmological dark matter. According to the numerous abservational
data:
\be
\Omega_{DM} = 0.26\,\,\, {\rm and}\,\,\,\, \Omega_{CMB} = 5.5\cdot 10^{-5}
\label{Omega-DM}
\ee
 or $\left(\rho_X / \rho_\gamma \right)_{obs} \approx 5\cdot 10^3$. 
 }
 
 {
 As calculated e.g. in the book~\cite{KAB}, 
 the frozen cosmological mass density of $X$-particles is determined by the equation:
  \be
 \Omega_X h^2 \approx \frac{10^9 x_f}{m_{Pl}\, {\rm GeV} \,(\sigma_{ann}v)} \approx 0.12,
 \label{Omega-h2}
 \ee
where $h \approx 0.67 $ is the dimensionless Hubble parameter and $x_f = T_f /m_X= 20-30$
is the ratio of the freezing temperature to the $X$ mass. The last term in the equation above 
is the observed value. Hence
\be
\sigma_{ann} v m_{Pl} \sim 3\cdot 10^{11} \cdot{\rm GeV}^{-1}
\label{sigm-m-Pl}
\ee
 and
 \be 
 n_X \approx 10^{-12} z^{-3}T_{in}^2 \cdot {\rm GeV}. 
 \label{nX-fin}
 \ee
 So for the ratio of X to relativistic particles densities we find:
  \be
 \frac{n_X}{n_{rel}} \rightarrow 10^{-12}  {\rm GeV}/T_{in} \approx 7\cdot 10^{-12}. 
 \label{final-ratio}
 \ee
 and the ratio of the corresponding energy densities at the present time
 \be
 \frac{\rho_X}{\rho_{CMB}} = \frac{m_X}{3 T_{CMB}} \frac{n_X}{n_{rel}}
 \frac{g_* (0.1{\rm MeV)}}{g_* (150){\rm MeV}} < 10^{3} \frac{m_X}{\rm Gev},
 \label{rat-rho}
 \ee
 which is safely below the observer ratio ${\rho_X}/{\rho_{CMB}} = 5\cdot 10^3$,
 especially if $m_X < 1 $ TeV.
 Here we took $g_* = 50$ at $T= 150$ MeV and $g_* = 1$ at $T=0.1 $ MeV.
 }
 
 One can see that the results presented in this Appendix disagree with the published
works~\cite{fujita} and ~\cite{lennon} on production of possible dark matter particles by PBH evaporation. But the
disagreement is natural, since in these papers some essential physical effects are disregarded. 
Firstly, it is assumed that the evaporation goes into an empty space, while in our case the universe was 
filled by cooler relativistic plasma. Secondly, the residual annihilation of the created DM particles is disregarded, 
while as it is shown above it is very much essential. The cooling of DM particles is so fast that their inverse
annihilation does not take place.

\section{Appendix B}
We present here  analytic expressions for the integrals of  $I_0$ (\ref{I-0}) and $I_3$ (\ref{I-3})  for two forms of PBH mass
spectrum: flat one and (the first index of $j$ is 1) and
that numerically close to the log-normal one  (the first index of $j$ is 2), see eq. (\ref{f-2}) and above. 
The second indices 1 or 3 correspond $I_0$ and $I_3$ respectively.
For brevity we use notations $t$ instead of $\eta$.

\begin{figure}[!hbtp]
\centering
\includegraphics[
width=0.9\textwidth]{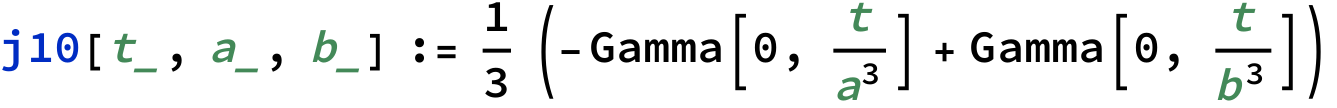}
\caption{The analytic result for the integral $j_{10} $ defined in eq.~(\ref{int-F1-0})
}
\label{int-anal-2}
\end{figure} 

\begin{figure}[!hbtp]
\centering
\includegraphics[
width=0.65\textwidth]{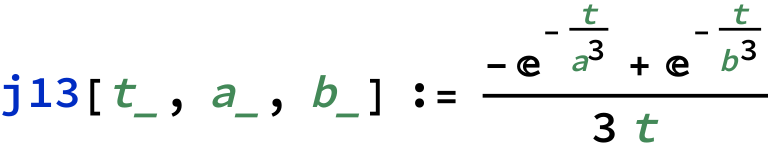}
\caption{The analytic result for the integral $j_{13} $ defined in eq.~(\ref{int-F1-3})
}
\label{j13-anal}
\end{figure}

\begin{figure}[!hbtp]
\centering
\includegraphics[
width=1.1\textwidth]{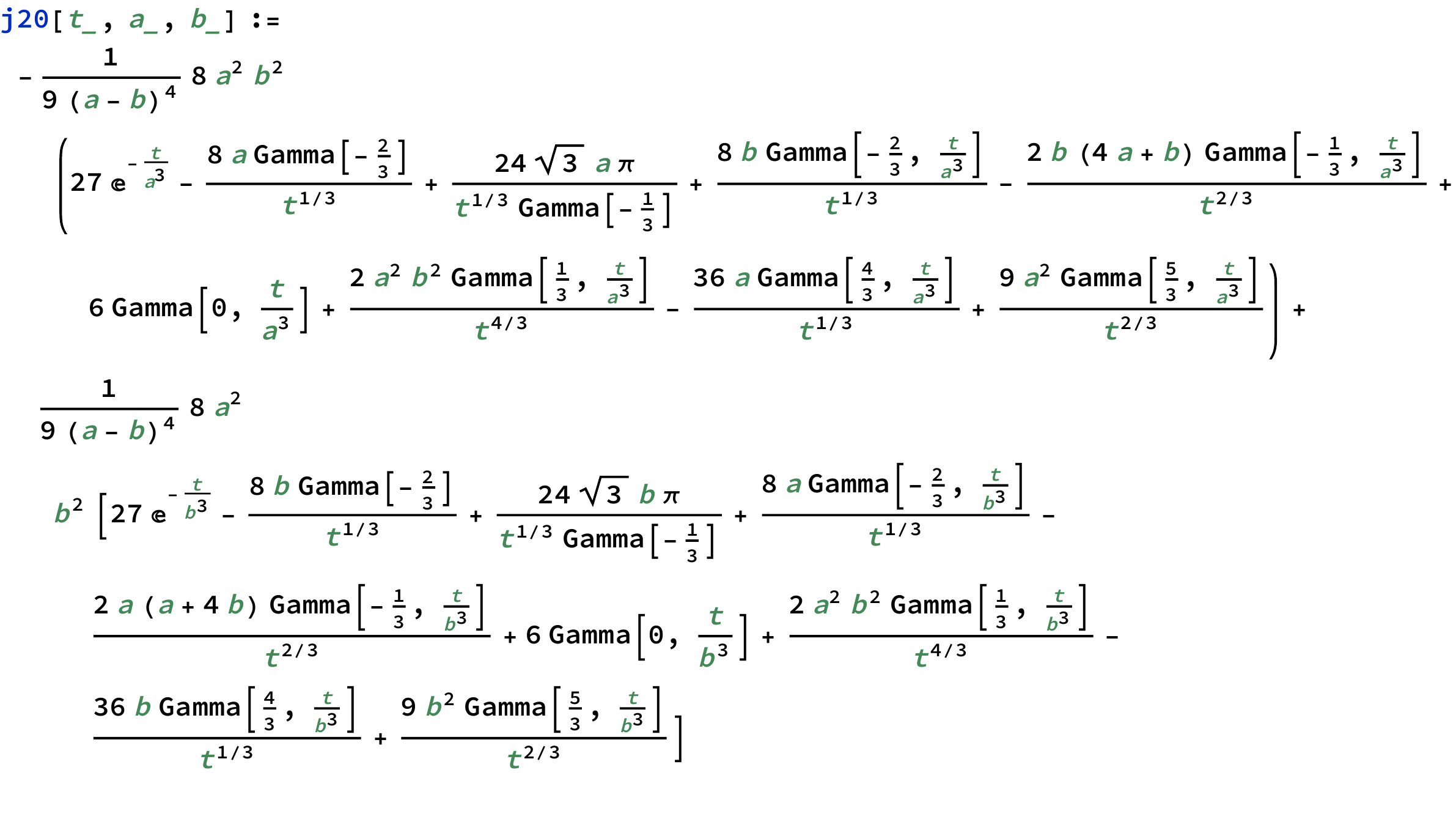}
\caption{The analytic result for the integral $j_{20} $ as explained in subsection~\ref{ss-log-norm}
}
\label{j20-anal}
\end{figure}

\begin{figure}[!hbtp]
\centering
\includegraphics[
width=1.1\textwidth]{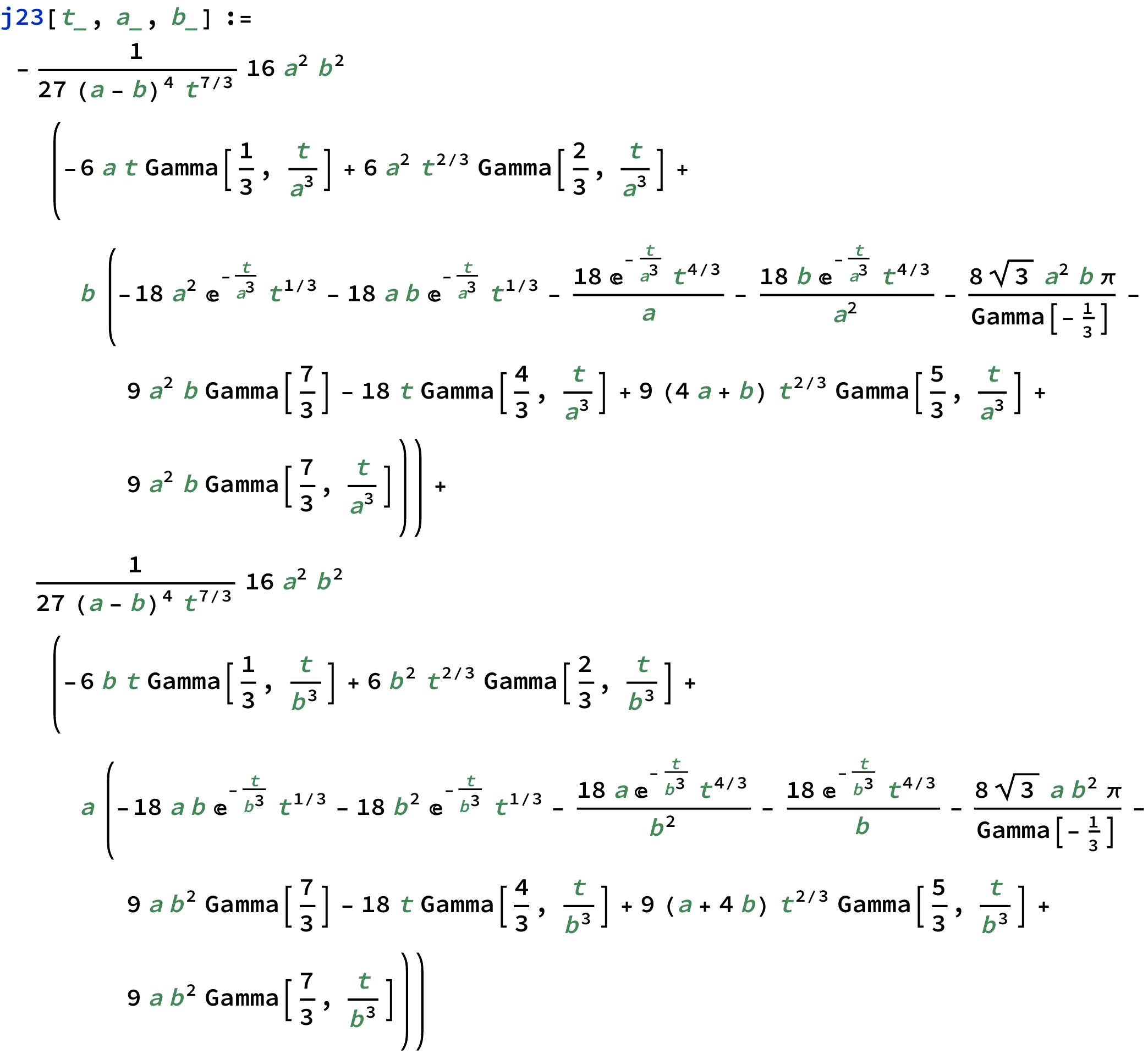}
\caption{The analytic result for the integral $j_{23} $ as explained in subsection~\ref{ss-log-norm}
}
\label{j23-anal}
\end{figure}


\end{document}